\documentclass[referee]{aa}

\newcommand{\unit}[1]{\; \mbox{#1}}

\usepackage{amssymb}
\usepackage{graphicx}
\usepackage{aas_macros,natbib}

\begin{document}

\title{Spatial Origin of Galactic Cosmic Rays in Diffusion Models: 
I- Standard sources in the Galactic disk}
\author{R.Taillet\inst{1,2} \and D. Maurin\inst{1,3}}

\authorrunning{Taillet \& Maurin}
\titlerunning{Spatial Origin of Cosmic Rays in Diffusion Models}
\institute{Laboratoire de Physique Th\'eorique {\sc lapth},
    Annecy--le--Vieux, 74941, France
    \and
    Universit\'e de Savoie, Chamb\'ery, 73011, France
    \and
    Institut d'Astrophysique de Paris, 98 bis Bd Arago, 75014 Paris, France
}

\date{Received \today; accepted}

\abstract{
The propagation of Galactic Cosmic Ray nuclei having energies between 100
MeV/nuc and several PeV/nuc is strongly believed to be of
diffusive nature. The particles emitted by a source located in the disk 
do not pervade the whole Galaxy, but are rather
confined to a smaller region whose spatial extension is related to
the height of
the diffusive halo, the Galactic wind and the spallation rate.
Following the pioneering work of \cite{Jones78}, this paper presents 
a general study on the {\em spatial} origin of cosmic rays,
with a particular attention to the role
of spallations and Galactic wind.
This question is different, and to a certain extent disconnected, 
from that of the {\em origin} of cosmic rays.
We find the regions of the disk from which a given 
fraction of cosmic rays detected in the Solar neighborhood were emitted
($f$-surfaces). After a general study, we apply the results to a
realistic source distribution, with the
propagation parameters obtained in our previous systematic
analysis of the observed secondary-to-primary ratios \citep{Maurin02b}.
The shape and size of these $f$-surfaces depend
on the species as well as on the values of the propagation parameters.
For some of the models preferred by our previous 
analysis (i.e. large diffusion slope $\delta$), 
these $f$-surfaces are small and in some extreme cases
only a fraction of a percent of the whole Galactic sources actually contribute 
to the Solar neighborhood Cosmic Ray flux.
Moreover, a very small number of sources may be responsible for more than 15 \% of the 
flux detected in the Solar neighborhood.
This may point towards the necessity to go beyond the
approximations of both homogeneity and stationarity.
Finally, the observed primary composition
is dominated by sources within a few kpc.
}

\maketitle

%%%%%%%%%%%%%%%%%%%%%%%%%%%%%%%%%%%%%%%%%%%%%%%%%%%%%%%%%%%%%%%%%%%%%%%%%%%%%%%%%%%%
%%%%%%%%%%%%%%%%%%%%%%%%%%%%%%%%%%%%%%%%%%%%%%%%%%%%%%%%%%%%%%%%%%%%%%%%%%%%%%%%%%%%

\section{Introduction}

The propagation of charged Cosmic Ray nuclei, in the energy range going
from a few 100 MeV/nuc and a few PeV/nuc, is strongly affected by the 
Galactic magnetic field. It is a diffusive process, so that the 
Cosmic Rays emitted by a single source  spread out in time,
pervade the whole Galaxy, and can escape the Galaxy when reaching 
its boundaries.
Those coming from a source located far from the Sun have a
larger probability of escaping than reaching the Solar neighborhood.
It is the opposite for nearby sources, so that the 
cosmic ray fluxes in the solar neighborhood are more sensitive 
to the properties of the local sources (as opposed to the remote sources).
Other effects like spallations and Galactic wind further limit the
distance Cosmic Rays travel before being detected.
Some consequences of the Galactic wind were studied in
\citet{Jones78} where convective escape was compared to
escape through the top and bottom boundaries of the Galaxy.

The goal of this paper is to go one step beyond by 
providing a general study on 
the {\em spatial} origin of cosmic rays, i.e. to answer the question 
"from which region of the Galaxy were emitted the cosmic rays detected in the 
solar neighborhood?".
This question is different, and to a certain extent disconnected, 
from that of the {\em origin} of cosmic rays ("What 
are the astrophysical objects which are responsible for the 
acceleration of cosmic rays?") which is still much debated.
We believe that it is nevertheless an interesting question, for 
several reasons. First, we find that the answer may cast some doubt 
on the validity of the stationary model, upon which most studies on 
Cosmic Rays are based. Second, it gives some clues about the spatial range 
beyond which the cosmic ray studies are blind to the sources.
Finally, this study may be of interest to optimize the propagation 
codes based on Monte-Carlo methods, by focusing the numerical effort on 
the sources that really contribute to the detected flux.

The reader who does not want to go through the pedagogical
progression can go directly from the general presentation of the method
in $\S$~\ref{general_formula} to its application in realistic cases in
$\S$~\ref{BC_induced}.
For the others, the effect of escape is studied in $\S$~\ref{bound}
and that of spallations and Galactic wind is studied in 
$\S$~\ref{spal_et_conv}.
Then, $\S$~\ref{resultats_de_base} studies the effect of a realistic
source distribution.
Finally, the fully realistic case is considered in $\S$~\ref{BC_induced}.
The results and the perspectives are discussed in the last section.
For convenience, we will use the word $f$-surfaces
to describe the surfaces in the thin disk within which the sources 
form the fraction $f$ of cosmic rays detected at the observer location.

%%%%%%%%%%%%%%%%%%%%%%%%%%%%%%%%%%%%%%%%%%%%%%%%%%%%%%%%%%%%%%%%%%%%%%%%%%%%%%%%%%%%
%%%%%%%%%%%%%%%%%%%%%%%%%%%%%%%%%%%%%%%%%%%%%%%%%%%%%%%%%%%%%%%%%%%%%%%%%%%%%%%%%%%%

\section{Description of the method}
\label{general_formula}

A stationary point source emits particles that diffuse in a given
volume. At the boundaries of this volume, the particles are free to
escape and the density drops to zero.
After a sufficiently long time, the stationary regime is eventually
reached and the density profile is established inside the diffusive
volume. If several sources are present (or even a continuous
distribution of sources), their contributions add linearly at each
point.

The question we wish to answer is the following:
a Cosmic Ray being detected at the position $\vec{r}_o$ of an observer
(in practice, this will be the position of the Sun, and we refer
to this position as the {\em Solar neighborhood}\/), what is the
probability density
\begin{equation}
         \frac{d{\cal P} \left\{{\rm emitted:~} \vec{r}_s,\;\vec{r}_s+d\vec{r}_s
         ~|~ {\rm observed:~} \vec{r}_o\right\}}{d\vec{r}_s} \; \equiv
         \frac{d{\cal P} \left\{ \vec{r}_s |  \vec{r}_o  \right\}}{d\vec{r}_s}
\end{equation}
that this Cosmic Ray was emitted from a source located at the position
$\vec{r}_s$~?
Such a question falls among classical problems of statistics.
A rigorous theoretical frame is provided by the Bayes approach that
summarizes the
proper use of conditional probabilities. A cruder but sufficient (and
equivalent) treatment is given by the frequency interpretation.
The probability written above is simply given by
\begin{equation}
          \frac{d{\cal P} \left\{ \vec{r}_s |  \vec{r}_o  \right\}}{d\vec{r}_s}
         = \frac{d{\cal N} \left[ \vec{r}_s \rightarrow \vec{r}_o 
\right]/d \vec{r}_s}
         {{\cal N} \left[ \rightarrow \vec{r}_o \right] }\;\;,
\end{equation}
where ${\cal N} \left[ \rightarrow \vec{r}_o \right]$ is
the number of paths reaching $\vec{r}_o$ and
$d{\cal N} \left[ \vec{r}_s\rightarrow \vec{r}_o \right]/d\vec{r}_s $ is
the density of paths going from $\vec{r}_s $ to $\vec{r}_o$.
We finally notice that the latter number determines the
density of Cosmic Rays that reach the position $\vec{r}_o$, when a
source is placed at $\vec{r}_s$.
We can thus write
\begin{equation}
    \frac{d{\cal P}\left\{ \vec{r}_s |  \vec{r}_o  \right\}}{d\vec{r}_s}
    \propto \frac{d{\cal N} \left[ \vec{r}_s\rightarrow \vec{r}_o
    \right]}{d\vec{r}_s}
    \equiv N_{\rm r_s}(\vec{r}_o)\;\;,
    \label{proba_elementaire}
\end{equation}
where the density $N_{\rm r_s}(\vec{r}_o)$ is the solution of the
propagation equation for a point source located at $\vec{r}_s$.
The normalization factor in this  relation is obtained by imposing
that $d{\cal P}/d{\vec{r}_s}$ actually is a probability density, i.e. is
normalized to unity. We refer to the contours on which the
probability density is constant as {\em isodensity contours\/}.

If the sources are distributed according to $w(\vec{r}_s)$,
the probability that a Cosmic Ray detected at $\vec{r}_o$ was emitted
from a surface ${\cal S}$ is given by
\begin{equation}
      {\cal P} \left\{ {\cal S} | \vec{r}_o \right\}=
      \frac{\int_{\cal S} w(\vec{r}_s) N_{\rm r_s}(\vec{r}_o) d\vec{r}_s}
      {\int_{\cal S_{\rm tot}} w(\vec{r}_s) N_{\rm r_s}(\vec{r}_o)
      d\vec{r}_s}\;\;.
      \label{proba_integree}
\end{equation}
This probability contains all the physical information about the
spatial origin of Cosmic Rays. We define the $f$-surfaces, inside which the sources
contribute to the fraction $f$ of the detected flux, by the relation 
${\cal P} \left\{ {\cal S} | \vec{r}_o \right\}=f$.
Actually, even for a given value of $f$, there are many different surfaces, 
delimited by different closed contours, fulfilling this condition.
We focus on the smallest of these surfaces, which is precisely delimited 
by an isodensity contour.
We also use the term $r_{\rm lim}$-probability for the quantity
${\cal P} \left\{ r_s < r_{\rm lim} | \vec{r}_o  \right\}$.

%%%%%%%%%%%%%%%%%%%%%%%%%%%%%%%%%%%%%%%%%%%%%%%%%%%%%%%%%%%%%%%%%%%%%%%%%%%%%%%%%%%%
%%%%%%%%%%%%%%%%%%%%%%%%%%%%%%%%%%%%%%%%%%%%%%%%%%%%%%%%%%%%%%%%%%%%%%%%%%%%%%%%%%%%

\section{The escape through the diffusive volume boundaries}
\label{bound}

The region in which diffusion occurs is limited by surfaces
(hereafter the {\em boundaries}\/) beyond which
diffusion becomes inefficient at trapping the particles, so that
they can freely
escape at a velocity close to $c$.
The density outside the diffusive volume is very small, and it is
very reasonable to suppose that the boundaries are {\em absorbers\/}, i.e.
they impose a null density ($N=0$).

It is well-known that the shape and location of the boundaries play a crucial
role for diffusive propagation.
This section shows that the Cosmic Rays emitted from standard
sources in the disk are not sensitive to the radial extension of the
Galaxy, but only to its top and  bottom edge.
To this aim, it is sufficient to concentrate on pure diffusion and to neglect
spallations, the Galactic wind and reacceleration.
Indeed this is a conservative case as these effects
can only make the diffusion process even {\em less\/} sensitive to the presence
of the boundaries  (see below). Moreover, we consider the case of
a homogeneous source distribution located in the disk
$w(\vec{r}_s)\propto \delta(z)$, which also leads to a conservative result
if compared to a realistic radial distribution of sources.

We first consider the pure diffusion
equation with a Dirac source term 
\begin{equation}
-K\triangle N(\vec{r})=\delta(\vec{\vec{r}-\vec{r}_s})\;\;.
\label{pure_diff}
\end{equation}
In unbounded space, the solution is given by
$N_{\rm r_s}(r_o) =1/4\pi K ||\vec{r_o}-\vec{r_s}||$.
The influence of the boundaries is estimated by solving this equation
in three situations: first we consider only a side boundary, then
only a top plus bottom boundary, and finally all the boundaries.

                           %---------------------%

                           %---------------------%

\subsection{Boundaries influence}
\label{sec:bound}

\begin{figure}[ht]
      \centerline{\includegraphics*[width=0.5\columnwidth]{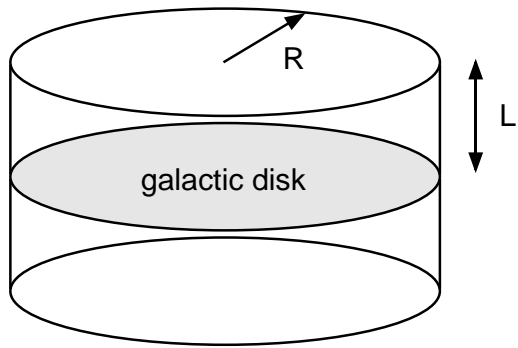}}
      \caption{Geometry of the diffusive volume.}
      \label{fig:dessin1}
\end{figure}
Our Galaxy can be represented as a cylindrical box with radial
extension $R$ and height $L$ (see App.~A for further details).
The probability density $d{\cal P}_{\rm cyl}(\vec{r}_s|\vec{r}_o)/d\vec{r}_s$
can be computed for arbitrary source and observer positions
$\vec{r}_s$ and $\vec{r}_o$,
using a Fourier-Bessel decomposition of the density.
In our case, the observer is located near the Sun, at a Galactocentric
distance $R_\odot \sim 8.5$~kpc. Unless the diffusive halo height is very
large, the top and bottom boundaries located at $z=\pm L$ are nearer
to us than the side boundary located at $R=20$~kpc.
As a result, we expect the effect of the side boundary to be smaller.
The first simplified situation we consider is that of an
observer located at the center of the Galaxy.
(In the case of an infinite disk, i.e. $R\rightarrow\infty$, this amounts
to a mere redefinition of the origin of the disk).

With $\vec{r}_o = \vec{0}$, the solution for a point source in this particular
geometry is given in App.~A.
The probability density that a particle reaching the observer was
emitted from a point located at a distance $r_s$ from the center is
thus given by (with $\rho_s\equiv r_s/R$)
\begin{equation}
       d{\cal P}_{\rm cyl}(\vec{r}_s|O)=\frac{d^2\vec{r}_s}{2\pi R^2} \times
       \left\{
       \sum_{i=1}^\infty  \frac{J_0(\zeta_i \rho_s)}
       {\zeta_i J_1^2(\zeta_i)} \times \tanh(\zeta_iL/R)
       \right\} \; . \;
       \left\{
       \sum_{i=1}^\infty  \frac{\tanh(\zeta_iL/R)}
       {\zeta_i^2 J_1(\zeta_i)}
       \right\}^{-1}\;\;,
       \label{infl_L2}
\end{equation}
normalization being obtained by imposing $\int_{0}^R
d{\cal P}(r_s|O)= 1$. The $r_{\rm lim}$-probability is given by
\begin{equation}
      {\cal P}_{\rm cyl} (r_s < r_{\rm lim}|  O  ) =
      \frac{r_{\rm lim}}{R}
      \left\{
      \sum_{i=1}^\infty  \frac{J_1(\zeta_i r_{\rm lim}/R)}
      {\zeta_i^2 J_1^2(\zeta_i)} \times \tanh(\zeta_iL/R)
      \right\} \; . \;
      \left\{
      \sum_{i=1}^\infty  \frac{\tanh(\zeta_iL/R)}
      {\zeta_i^2 J_1(\zeta_i)}
      \right\}^{-1}\;.
      \label{pourcent_cyl}
\end{equation}
This probability is independent of the value of the diffusion
coefficient $K$.

                           %*****%

\subsubsection{Side boundary}

\begin{figure}[ht]
      \centerline{\includegraphics*[width=0.5\columnwidth]{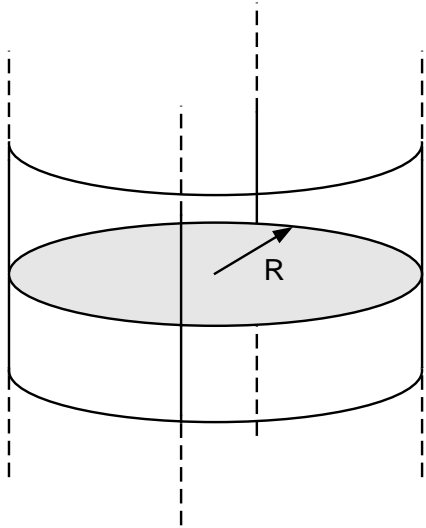}}
      \caption{Geometry of the diffusive volume in the limit
      $L\rightarrow \infty$.}
      \label{fig:dessin2}
\end{figure}
The escape from the side 
boundary (located at $r=R$) is disentangled from the escape 
from the $z=\pm L$ boundary by first considering the
limit $L \rightarrow \infty$.
For the sake of simplicity, we will, as above, only study the effect
of this boundary on
observations performed at the center of the Galaxy.
In the limit  $L \rightarrow \infty$, we have
$\coth(\zeta_iL/R)\approx 1$ in expression~(\ref{infl_L2}).
This gives for the $r_{\rm lim}$-probability,
\begin{displaymath}
    {\cal P}_{\rm R} (r_s < r_{\rm lim}|  O  ) =
    \frac{r_{\rm lim}}{R}
    \left\{
    \sum_{i=1}^\infty  \frac{J_1(\zeta_i r_{\rm lim}/R)}
    {\zeta_i^2 J_1^2(\zeta_i)}
    \right\} \; . \;
       \left\{
       \sum_{i=1}^\infty  \frac{1}
       {\zeta_i^2 J_1(\zeta_i)}
       \right\}^{-1}\;\;.
\end{displaymath}

                           %*****%

\subsubsection{Top and bottom boundaries}
\label{elec_image}

\begin{figure}[ht]
      \centerline{\includegraphics*[width=0.5\columnwidth]{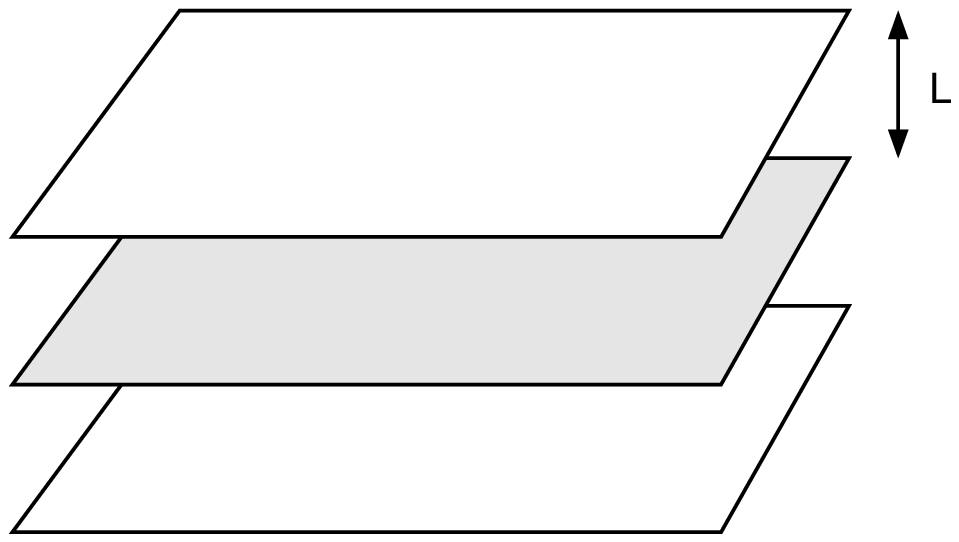}}
      \caption{Geometry of the diffusive volume in the limit
      $R\rightarrow \infty$.}
      \label{fig:dessin3}
\end{figure}
The influence of the $z=\pm L$ boundaries, in the case
of an infinite disk ($R \rightarrow \infty$) is now considered.
In this limit, the sum over Bessel functions can be replaced by an
integral and we obtain (see App.~B.3)
\begin{equation}
    d{\cal P}_{\rm L}(\vec{r}_s|O) \propto
    \frac{d^2\vec{r}_s}{r_s} \, \int_0^\infty J_0(x)
    \, \tanh \left( \frac{xL}{r_s} \right) \, dx\;,
    \label{infl_bords_int}
\end{equation}
which allows to compute the $r_{\rm lim}$-probability 
${\cal P}_{\rm L} (r_s < r_{\rm lim}|  O  )$ as before,
which is a function of $r_{\rm lim}/L$ only.
These integrals are somewhat intricate to compute numerically, due to the
very slow convergence. 
In this particular case, the accuracy of the numerical calculation
can be checked
for $r \gg L$, as a detailed study of the function
(\ref{infl_bords_int}) shows that in
this limit
\begin{equation}
    \label{space_cowboy}
    N^{\rm L}_{(r_s,0)}(O) \approx \frac{1}{4\pi K r_s} \times 
    2\sqrt{\frac{r_s}{L}}
    \, e^{-\pi r_s/2L}\;.
\end{equation}
It is also noticeable that the quantity
\begin{equation}
       f_{\rm esc}(r_s) \equiv 1- \frac{N^L}{N^{L=\infty}}
       = \int_0^\infty J_0(x) \, \left\{ 1-\tanh \left( \frac{xL}{r_s} \right)
       \right\}\, dx
\end{equation}
gives the fraction of Cosmic Rays emitted from a distance $r_s$ that
has escaped the diffusive halo before reaching us.

%---------------------%

\subsection{Summary: the effect of boundaries on primary species}

\begin{figure}[ht]
      \centerline{\includegraphics*[width=0.8\columnwidth]{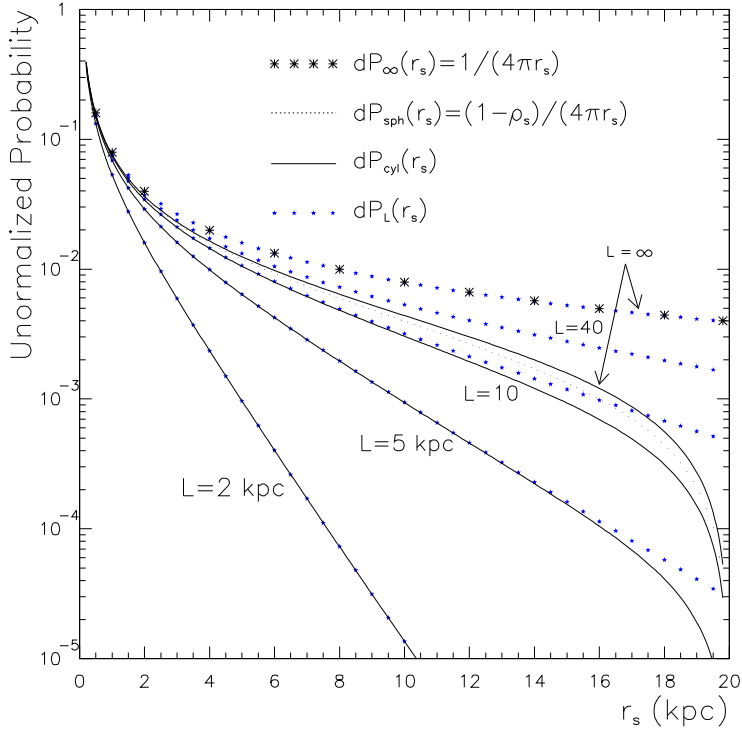}}
      \caption{Cosmic ray probability density as a function of
      $r_s$ (distance of the $\delta(\vec{r_s})$ source in the disk), for
      several values of $L$ and for a disk of radius $R=20$ kpc.
      Big stars are for unbounded model, dotted line is for a spherical
      boundary at radius $R$,
      small stars are for top and bottom boundaries, and
      solid lines are for cylindrical boundaries.}
      \label{fig:influence_r1}
\end{figure}

Fig.~\ref{fig:influence_r1} shows the probability
density computed above as a function of $r_s$ for unbounded space,
for the cylindrical geometry with several halo
sizes $L$, i.e. Eq.~(\ref{pourcent_cyl}), and for the two limiting cases
corresponding to $L\rightarrow\infty$ or  $R\rightarrow\infty$.
We also show, in Tab.~\ref{table:blabla}, the radii of the 
$f$-surfaces, in the two cases $R= 20$~kpc and $R \rightarrow \infty$.
It can be noticed that even if the source distribution is infinite in
extent, the finite size of the halo limits the quantity of Cosmic Rays that
reach a given point.
The mean distance from which the Cosmic Rays reaching the center
is given by $\langle r_s \rangle = 1.4 \, L$.
This effect
dominates over the leakage through the side boundary $r_s=R$,
and it will be even more negligible in realistic
situations, as (i) the source density is small near the edge of the
disk, (ii) when the spallations and Galactic wind are considered,
most Cosmic Rays are destroyed or blown out of the disk before they
have a chance to reach this side boundary.
\begin{table}[ht]
\begin{tabular}{|c|cc|cc|cc|}   \hline
        & \multicolumn{2}{c|}{$f(r_{\rm lim}) = 50$\%}  &
        \multicolumn{2}{c|}{$f(r_{\rm lim}) = 90$\%}
        & \multicolumn{2}{c|}{$f(r_{\rm lim}) = 99$\%}\\
        \hfill
        & $R=\infty$  & $R=20$~kpc & $R=\infty$ & $R=20$~kpc& $R=\infty$
&$R=20$~kpc\\
        \hline
         $L = \infty$ & -- & 6.2 kpc& -- & 14.1 kpc& -- & 18.2 kpc\\
         $L = 20$ &  12.6 kpc& 6.1 kpc&  39 kpc& 14 kpc&  68 kpc& 18.2 kpc\\
         $L = 5$ kpc &  3.1 kpc&  2.95 kpc &  9.5 kpc& 8.6 kpc& 17 kpc&
14.6 kpc\\
         $L = 1$ kpc &  0.63 kpc& 0.63 kpc&  1.9 kpc& 1.9 kpc& 3.4 
kpc& 3.4 kpc\\
         \hline
      \end{tabular}
      \caption{This table indicates the radius $r_{\rm lim}$
      inside which a given fraction $f(r_{\rm lim})\equiv
      {\cal P}(r_s < r_{\rm lim}|  O  )$ of Cosmic Rays reaching
      the center were emitted from, for several $L$ and in the case of the
      infinite disk and $R=20$~kpc.}
\label{table:blabla}
\end{table}

An important consequence is that as long as the observer and the
sources are not too close to the side boundary,
the density only depends on the relative distance
to the source in the disk, so that it may be assumed, for numerical
convenience, that the observer is either at the center of a finite disk,
or in an infinite disk.
In all the paper, i.e. for standard sources in the disk,
we will consider the limit $R \rightarrow \infty$,  i.e.
we use the integral representation described in App.~B.3.

%%%%%%%%%%%%%%%%%%%%%%%%%%%%%%%%%%%%%%%%%%%%%%%%%%%%%%%%%%%%%%%%%%%%%%%%%%%%%%%%%%%%
%%%%%%%%%%%%%%%%%%%%%%%%%%%%%%%%%%%%%%%%%%%%%%%%%%%%%%%%%%%%%%%%%%%%%%%%%%%%%%%%%%%%
\section{Secondary and radioactive species}

                          %---------------------%

\subsection{Progenitors of stable secondaries}
\label{progenitors}
As can be seen in App.~A.3, the secondary distribution
from point-like primary sources is related very simply to the primary
distribution itself. One could find strange to speak about
secondaries as we have not, for the moment, included spallations in the
model.
The right picture is the following:
a primary emitted at $r_s$ propagates and from time to time
crosses the disk (mostly filled with hydrogen, density $n_{\rm ISM}$).
During this crossing, there is
a probability $n_{\rm ISM}.v.\sigma_{{\rm prim}\rightarrow{\rm sec}}$
to create a secondary, that in turn propagates in the diffusive volume
until it reaches (or not) the experimental setup.
This will be taken into account properly in the next section.
However, in order to have a compact expression, a
crude estimation can be obtained by neglecting the influence of
spallations on the primary and secondary component.
This is obtained if one discards $\Gamma_{\rm inel}$ in the terms
$A_i^{\rm prim}$ and $A_i^{\rm sec}$ of Eq.~(\ref{sec.}).
The net result will be an overestimation
of the distance the secondaries come from since their destruction is
discarded two times; once under their primeval primary form and once
in their secondary form.

We find, in the case $R\rightarrow \infty$ (see App.~B.3), and for a
homogeneous distribution of sources,
\begin{displaymath}
         \frac{d{\cal P}_{\rm sec}(r_s|R_\odot)}{d^2\vec{r}_s} \propto 
\int_0^\infty  \frac{J_0(x)}{x}
         \times \tanh^2 \left( \frac{x L}{r_s} \right) \, dx\;.
\end{displaymath}
The resulting integrated probabilities are shown in
Tab.~\ref{table_origine_sec}.
The source of the primary that will give the secondaries observed at
a given point is located farther away than the sources of the
primary we detect (compare Tabs.~\ref{table_origine_sec} and
\ref{table:blabla}).
This may be of importance if for instance the source composition or
the source intensity varies with position: in the ubiquitous secondary-to-primary
ratio, the numerator is sensitive to sources located on a greater
range than the denominator.
Moreover, these secondaries set the size of an effective ``local" zone
outside of which the particles reaching the Solar neighborhood have
never been.
The local observations tell nothing about the propagation conditions
outside of this zone. 
One could object that this conclusion is
mainly based on the $f$-surfaces which refer to the
sources contributing to observed CR, but that the Cosmic Rays
reaching us from these sources actually sample (via random
walk) a much larger volume. This is actually not the case, as a 
particle wandering too far has a very small probability to 
ever come back to us. This point can be made more quantitative, as a 
simple reasoning shows that the probability ${\cal P}[ACB]$ that a 
particle emitted 
in $A$ and reaching $B$ has passed through $C$ is given by
${\cal P}[AC]{\cal P}[CB]$, which is closely related to 
$N_A(\vec{r}_C) \times N_B(\vec{r}_C)$. This later quantity is small 
as soon as $C$ is too far from $A$ or $B$.
\begin{table}[ht]
\begin{tabular}{|c|cc|cc|cc|}   \hline
        & \multicolumn{2}{c|}{$f(r_{\rm lim}) = 50$\%}  &
        \multicolumn{2}{c|}{$f(r_{\rm lim}) = 90$\%}
        & \multicolumn{2}{c|}{$f(r_{\rm lim}) = 99$\%}\\
        \hfill
        & $R=\infty$  & $R=20$~kpc & $R=\infty$ & $R=20$~kpc& $R=\infty$
&$R=20$~kpc\\
        \hline
         $L = \infty$ & -- & 8.6 kpc& -- & 15.3 kpc& -- & 18.5 kpc\\
         $L = 5$ kpc &   5.5 kpc&  5.3 kpc &  12.5 kpc& 12 kpc& 25 kpc&
17.2 kpc\\
         $L = 1$ kpc &  1.1 kpc& 1.1 kpc&  2.5 kpc& 2.5 kpc& 4.4 kpc& 4.4 kpc\\
         \hline
      \end{tabular}
\caption{This table indicates the radius inside which a given fraction
$f(r_{\rm lim})$ of secondary Cosmic Rays reaching the center were 
emitted from, for several
$L$ and in the case of a disk of radius $R=20$ kpc.
The last line shows that for small $L$, the effect of the side
boundary is completely negligible.}
\label{table_origine_sec}
\end{table}

                            %*****%

     \subsection{Radioactive secondaries}
     \label{partie_rad}
In the case of an unstable species  with a lifetime $\tau$, formula
(\ref{sol_prim}) can be written as
\begin{equation}
          d{\cal P}_{\rm rad} \left\{ r_s |  O  \right\}  \propto
          d\rho_s \, \sum_{i=1}^\infty  \frac{J_0(\zeta_i \rho_s) }{
          \sqrt{R^2 \Gamma_{\rm rad}/K + \zeta_i^2} \, J_1^2(\zeta_i)}\;,
          \label{rad}
\end{equation}
where $\Gamma_{\rm rad}= \tau^{-1} = \gamma^{-1}\tau_0^{-1} $.
This expression can be transformed using the identity \citep{Lebedev}
\begin{displaymath}
          \frac{1}{\sqrt{\zeta_i^2 + \alpha^2}} =
          \int_0^\infty \frac{e^{-\alpha \rho}}{\rho} \rho  J_0(\zeta_i
          \rho) d\rho
          \approx
          \int_0^1 \frac{e^{-\alpha \rho}}{\rho} \rho  J_0(\zeta_i
          \rho) d\rho\;.
\end{displaymath}
The approximation in the last step is valid
if the exponential term decreases with $\rho$ fast enough
(i.e. $\alpha$ is large so that the upper limit can be set to 1 in
the integral).
We then recognize in (\ref{rad}) the Fourier-Bessel transform of 
$\exp(-\alpha \rho)/\rho$, so that finally
the normalized probability reads
\begin{equation}
          d{\cal P}_{\rm rad} \left\{ r_s |  O  \right\} =
          \frac{\exp(-r_s/l_{\rm rad})}{2\pi \,r_s .\, l_{\rm rad}}
          \,  d^2\vec{r}_s\;,
          \label{rir4}
\end{equation}
where the following typical length has been introduced
\begin{equation}
      l_{\rm rad}= \sqrt{\frac{K}{\Gamma_{\rm rad}}} =
      0.17 \unit{kpc} \times \sqrt{\frac{K}{0.03
      \unit{kpc}^2\unit{Myr}^{-1}} }
      \sqrt{\frac{\tau}{ 1 \unit{Myr}} }\;.
      \label{f_rad}
\end{equation}
Indeed, this result can be derived much
more straightforwardly starting from the stationary equation
$ -K\Delta_r N(r) +\Gamma_{\rm rad}N(r)=0$ (with a source at the origin) in
unbounded space. This is also in full agreement with the expression given
in App.~B (see also Sec.~4.1) of \cite{Donato02},
where we found the same expression starting from the propagator of the
non-stationary diffusion equation in unbounded space.

To sum up, Eq.~(\ref{rir4}) is valid as long as $l_{\rm rad} \ll R$ and
$l_{\rm rad} \ll L$: the propagation of the unstable species can be then
considered as {\em local}\/, with a typical scale $l_{\rm rad}$.
This is no longer the case if the lifetime $\tau = \gamma \tau_0$ is large,
which is the case at high energy because of the relativistic factor
$\gamma$, even if the proper lifetime $\tau_0$ is short.
The $r_{\rm lim}$-probability is straightforwardly derived. 
As on these typical scales,
the source distribution can safely taken to be constant, the distance
$r_{\rm lim}$ is expressed as
\begin{equation}
r_{\rm lim}=-l_{\rm rad}\times \ln (1-f)\;.
\label{brungle2}
\end{equation}
It means that the sources that contribute to the fraction 
$f=(50-90-99)$\% of the radioactive
species measured flux are located inside the disk of radius
$r_{\rm lim}=(0.7-2.3-4.6)\times l_{\rm rad}$. The effect of a
local underdensity around the Sun is 
discussed later.

                           %---------------------%

\subsection{Electrons and positrons}

Cosmic ray sources also emit electrons and positrons.
In contrast with the nuclei, these particles are light, so that they
are subject to much stronger energy losses, due to synchrotron radiation
and inverse Compton.
This results in an effective lifetime given by (e.g. \cite{Aharonian95})
$\tau_{\rm loss} \sim 300 \unit{Myr} \times (1 \unit{GeV}/E)$.
The results given in the previous section on radioactive species can
be applied to this case, with a scale length
\begin{displaymath}
      r_{\rm loss} \sim 1 \unit{kpc} \times \sqrt{\frac{1 \unit{GeV}}{E}}
      \sqrt{\frac{K}{0.03 \unit{kpc}^2\unit{Myr}^{-1}}}\;.
\end{displaymath}
Formulae~(\ref{rir4}) and~(\ref{brungle2}) can be used with $l_{\rm
rad}\leftrightarrow r_{\rm loss}$.
This effect is discussed by \cite{Aharonian95} to show that a nearby
source may be necessary to explain the high energy electron flux observed
on the Solar neighborhood.

\subsection{Summary: pure diffusive regime, an upper limit}

The important conclusions at this point are that
i) most of the stable primary Cosmic Rays that reach the Solar neighborhood
were emitted from disk sources located within a distance of the order 
of $L$, such that the $R=20 \unit{kpc}$ boundary can reasonably be discarded
ii) the secondary species composition is determined by sources
located farther away than those determining the primary composition;
iii) radioactive species may come from very close if their
lifetime is so short that $\sqrt{K \gamma \tau_0} < L$, high energy
electrons and positrons definitely do.

These conclusions are expected to be stronger when spallations, Galactic
wind and a realistic source distribution are taken into account.
All these effects will limit even more the range that the particles
can travel before reaching the Solar neighborhood.

%%%%%%%%%%%%%%%%%%%%%%%%%%%%%%%%%%%%%%%%%%%%%%%%%%%%%%%%%%%%%%%%%%%%%%%%%%%%%%%%%%%%
%%%%%%%%%%%%%%%%%%%%%%%%%%%%%%%%%%%%%%%%%%%%%%%%%%%%%%%%%%%%%%%%%%%%%%%%%%%%%%%%%%%%

\section{The effects of spallation and convection}
\label{spal_et_conv}

                  %---------------------%

\subsection{Pure convection}

\begin{figure}[ht]
       \centerline{
       \includegraphics*[width=\columnwidth]{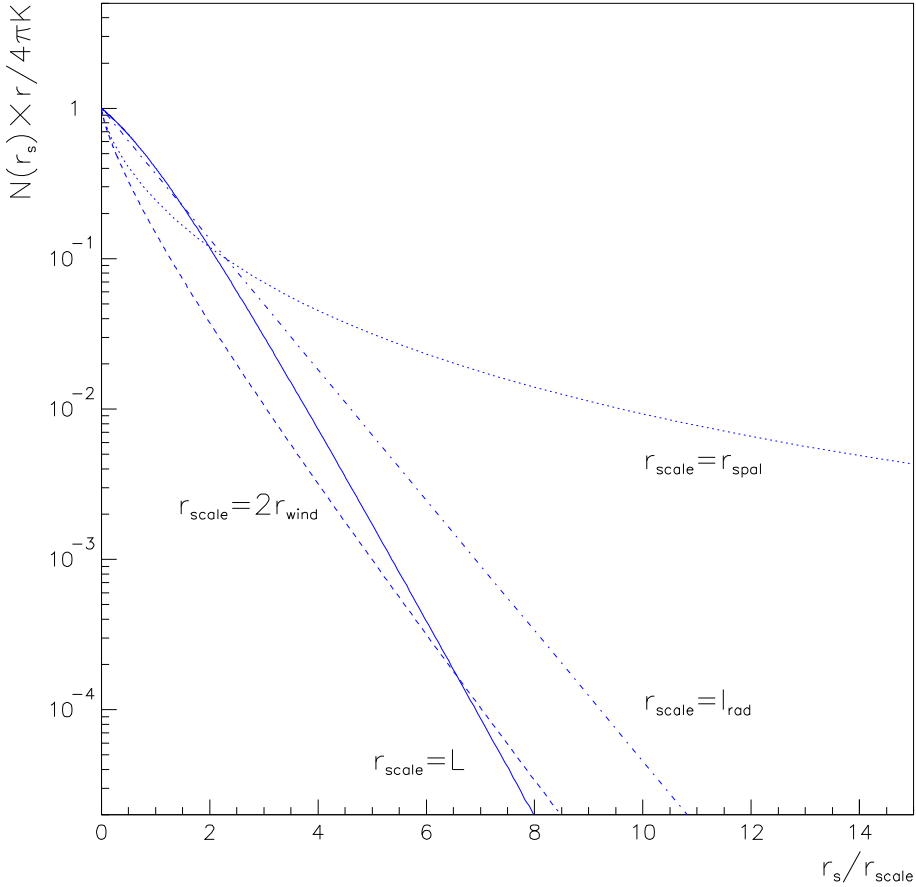}}
       \caption{Deviation from the pure $1/r_s$ density profile,
       $N(r_s)/(1/4\pi K r_s)$, due to the
       various effects studied here: escape from the $z=\pm L$
       boundaries, spallations and Galactic wind. In this latter case,
       the choice $r_{\rm scale}=2r_{\rm wind}$ has been made to show 
the similar
       behavior at large $r_s$.
       The case of a radioactive species has also been shown.
       It should be noticed, however, that in most interesting cases, the
       scale length $l_{\rm rad}$ is much smaller than the others, so that
       in this case the propagation is dominated by radioactive decay
       and spallations and Galactic wind can be safely discarded.}
       \label{fig:cutoff}
\end{figure}

The diffusion of Cosmic Rays may be disturbed by the presence of a
convective wind of magnitude $V_c$, directed outwards from the disk.
For numerical convenience, a constant wind has been considered, 
although other possibilities (especially a linear dependence) are probably more justified on 
theoretical grounds (see discussion in MTD02).
The effect is to blow the particles away from the disk, so that
those detected in the Solar neighborhood come from closer sources
(compared to the
no-wind case).
With an infinite halo, the probability density in the disk is
given by
\begin{eqnarray}
      \frac{d{\cal P}^{\rm wind}_{L \rightarrow
      \infty} \left\{r_s |  0  \right\}}{d^2\vec{r}_s}
      &\propto&  \int_0^\infty \frac{k J_0(kr_s)\, dk}{V_c + K
      \sqrt{
      V_c^2/K^2 + 4 k^2}}\nonumber\\
      &\propto& \frac{1}{r_s} \int_0^\infty \frac{x J_0(x)\, 
dx}{r_s/r_{\rm wind} +
      \sqrt{
      (r_s/r_{\rm wind})^{2} + x^2}}\;,
      \label{avec_vent}
\end{eqnarray}
where the characteristic radius $r_{\rm wind}\equiv 2K/V_c$ 
has been defined.
The expression in Eq.~(\ref{avec_vent}) is a function of $r_s/r_{\rm wind}$ only.
The deviation from a pure $1/r_s$ law, 
as well as deviations
due to escape, radioactive decay and spallation (see next section),
is shown in Fig.~\ref{fig:cutoff}.
The $r_{\rm lim}$-probability is given by
\begin{eqnarray}
       {\cal P}^{\rm wind}_{L \rightarrow\infty}(<r_{\rm lim})
       &=& \int_0^\infty \frac{J_0(x)}{4x}
       \times
       \left\{
       \frac{r_{\rm lim}}{r_{\rm wind}}
       \sqrt{\frac{r_{\rm lim}^2}{r^2_{\rm wind}}
       + x^2} - \frac{r_{\rm lim}^2}{r^2_{\rm wind}}\right.\nonumber\\
       &+& \left.x^2 \ln \left[\frac{r_{\rm lim}/r_{\rm wind} +
       \sqrt{r_{\rm lim}^2/r^2_{\rm wind}  +
       x^2}}{x}\right] \right\} \, dx\;.
\end{eqnarray}
Some values are indicated in Tab.~\ref{big_table} and
plotted in  Fig.~\ref{fig:influence_tot}.

It is interesting to note that the effect of $2r_{\rm wind}$ is similar (though
not rigorously identical) to the effect of $L$ (see Fig.~\ref{fig:cutoff}).
As a matter of fact, this was noticed by \cite{Jones78} who studied
the propagation properties in a dynamical halo and provided a very simple
picture (along with a rigorous derivation) of the effect of the wind.
Consider a particle initially located at a distance $z$ from the disk.
It takes a time $t_{\rm diff}\approx z^2/K$ to diffuse back in the disk.
In the meantime, convection sweeps the particle in a distance $z_w\equiv
V_c t_{\rm diff}\approx V_c z^2/K$. Both processes are in competition and
the particle will not reach the disk if $z_w>z$. This define an effective
halo size $L^*\approx K/V_c$. This is our parameter $r_{\rm wind}$ up to
a factor 2.
                         %---------------------%

\subsection{Pure spallation}

The Galactic disk contains interstellar gas mostly made of hydrogen.
When Cosmic Rays cross the disk, they can interact with this gas.
This interaction may result in a nuclear reaction (spallation),
leading to the destruction of the incoming particle and to the
creation of a different outgoing particle (secondary).
We present two approaches to the problem of diffusion in presence of
a spallative disk. When the halo is infinite in extent, the solution
may be obtained by using the interpretation of diffusion in terms of
random walks.
This will be treated in App.~C.
In the general case, the Bessel developments can be used as before.
Starting from Eq.~(\ref{sol_prim}), the expression for the
probability density is readily obtained.
The limit $L \rightarrow \infty$ is noteworthy, as the resulting
expression isolates the influence of spallations:
\begin{equation}
      \frac{d{\cal P}^{\rm spal}_{L \rightarrow\infty} \left\{ r_s | 
0  \right\}}{d^2\vec{r}_s}
      \propto    \int_0^\infty \frac{k J_0(kr_s)}{2h \Gamma_{\rm inel} 
+ 2kK} \, dk = \frac{1}{4\pi K r_s}
      \int_0^\infty \frac{x J_0(x)
      dx}{r_s/r_{\rm spal} + x}
      \label{expression_integrale_spal}
\end{equation}
where the quantity $r_{\rm spal}\equiv K/(h \Gamma_{\rm inel})$ 
has been defined.
Would there be no spallation, the $1/r_s$ behavior would be recovered.
The term $2h \Gamma_{\rm inel}$ has the effect to kill the
contributions of $k \lesssim k_{\rm spal}$ in the integral, with
$k_{\rm spal} \equiv h \Gamma_{\rm inel}/K$.
It leads to a decrease of the integral on scales $r>r_{\rm spal} =
1/k_{\rm spal}$.
Some typical values, for $K=\beta K_0 {\cal R}^\delta$
(see Sec.~\ref{subsec:energy})
with $K_0 = 0.03 \unit{kpc}^2 \unit{Myr}^{-1}$
and $\delta = 0.6$ are
given below at 1 GeV/nuc and 100 GeV/nuc.
\begin{table}[ht]
      \begin{tabular}{|c|ccc|}   \hline
         & p  & O  & Fe  \\
         \hline
         $\sigma~(\mbox{mb})$ & 44 & 309 & 760\\
         $r_{\rm spal}~(\mbox{kpc})$, 1 GeV/nuc & 10.2 & 1.45 & 0.59\\
         $r_{\rm spal}~(\mbox{kpc})$, 100 GeV/nuc & 115 & 16.4 & 6.7\\
         \hline
      \end{tabular}
      \caption{Some values of the inelastic cross section and the
      associated spallation scale length.}
\end{table}
The heavy species are more sensitive to spallations, so that they
come from a shorter distance.
This could in principle affect the mean atomic weight of Cosmic Rays
if the composition of the sources is not homogeneous
(see e.g. \cite{Maurin02a}). See Sec.~\ref{chi_wind_chi_spal}) for the results 
with realistic propagation parameters.

For small values of $r_s/r_{\rm spal}$, the convergence of the 
previous integral is
slow, and other forms obtained by integration by parts, as
developed in the App.~B.3, might be preferred.
However, in this particular case, the identity
\begin{displaymath}
    \int_0^\infty x dx J_0(x)/(x+\alpha)
    = \int_0^\infty y dy e^{-\alpha y}/(1+y^2)^{3/2}
\end{displaymath}
yields the more useful form
\begin{equation}
      \frac{d{\cal P}^{\rm spal}_{L \rightarrow\infty} \left\{ r_s |  0
      \right\}}{d^2\vec{r}_s}
      =   \frac{1}{4\pi K r_s}
      \int_0^\infty \frac{y e^{- y r_s/r_{\rm spal}}}{(1+y^2)^{3/2}} \; dy\;.
      \label{forme_utile}
\end{equation}
This expression is in full agreement with Eq.~(\ref{The_riri_equation})
obtained with the random walk approach (see App.~C).
For large values of $r_s/r_{\rm spal}$, the convergence can be checked
by comparing the results to the asymptotic development
\begin{displaymath}
     \int_0^\infty \frac{x J_0(x)}{x+\alpha} \; dx
     \approx \frac{1}{\alpha^2} - \frac{9}{\alpha^4}+
     \frac{225}{\alpha^6} + \ldots
\end{displaymath}

Finally, the $r_{\rm lim}$- probability can be computed as before
\begin{displaymath}
      {\cal P}^{\rm spal}_{L \rightarrow \infty}(<r_{\rm lim})
      =  \int_0^\infty \frac{dx}{\left(1+x^2\right) ^{3/2}} \,
      \left[ 1 - \exp\left\{-x\frac{r_{\rm lim}}{r_{\rm spal}} \right\}
      \right]\;.
\end{displaymath}
Some values are indicated in Tab.~\ref{big_table}.

%---------------------%
\subsection{Comparison and combination of the different effects}
\begin{figure}[ht]
      \centerline{
      \includegraphics*[width=\columnwidth]{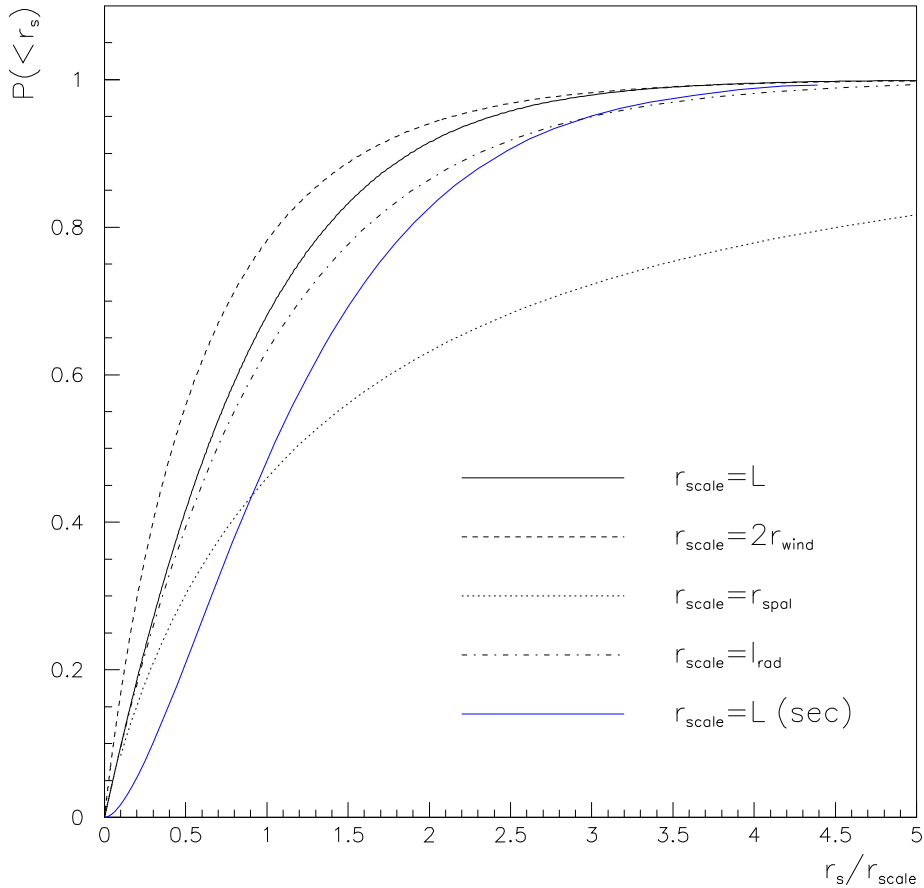}}
      \caption{Integrated probability that a particle detected at the origin
      was emitted inside the ring of radius $r_s$, in the three situations
      considered. The solid dark line is obtained when only the leakage through
      the $z=\pm L$ boundaries is considered, in which case the radii scale
      as $r/r_{\rm scale}$ with $r_{\rm scale} = L$.
      The dotted, respectively dashed, line is obtained when only the
      spallations, respectively only the convective wind, are considered.
      The solid grey line indicates the probability that the primary progenitor
      of a secondary detected in the Solar neighborhood was emitted
from within a given distance.
      }
      \label{fig:influence_tot}
\end{figure}

To summarize, the effect of spallation and Galactic wind
depends on the two parameters:
\begin{equation}
\left\{
\begin{array}{l}
\displaystyle
   r_{\rm wind} \equiv \frac{2K}{V_c} \approx
      5.87 \unit{kpc} \times\left(\frac{K(E)}{0.03
      \unit{kpc}^2\unit{Myr}^{-1}}\right)
      \left(\frac{ 10 \unit{km} \unit{s}^{-1}}{V_c}\right)\;\;;
      \vspace{0.2cm}\\\displaystyle
   r_{\rm spal}\equiv \frac{K}{h \Gamma_{\rm inel}} \approx
        3.17 \unit{kpc} \times \frac{1}{\beta} \times\left(\frac{K(E)}{0.03
        \unit{kpc}^2\unit{Myr}^{-1}}\right)
        \left(\frac{ 100 \unit{mb}}{\sigma}\right)\;\;.
\end{array}
\right.
\label{f_rspal}
\end{equation}
The $r_{\rm lim}$-probability is displayed in
Fig.~\ref{fig:influence_tot} as a function of $r_{\rm lim}$.
The effect of the Galactic wind is very similar to that of the top
and bottom boundaries, whereas the effect of spallations is quite
different. In the latter case, the cutoff in the density is a power
law in $r_s$ and decreases much more slowly than the exponential
cutoff due to the wind or to escape. As a result, the 99\%-surfaces
are much larger than the 90\%-surfaces.
This can also be seen in the first three lines of Tab.~\ref{big_table}.

When all the effects above are considered, Eq.(\ref{sol_prim}) gives
\begin{displaymath}
       \frac{d{\cal P} \left\{ r_s |  0  \right\}}{d^2\vec{r}_s}
      \propto \frac{1}{kr_s}
      \int_0^\infty \frac{x J_0(x)\, dx}
      {\rho_{\rm spal} + \rho_{\rm wind} + \sqrt{\rho_{\rm wind}^2 + x^2}
      \coth \left\{\sqrt{\rho_{\rm wind}^2 + x^2}/\rho_L\right\} }
\end{displaymath}
where $\rho_{\rm spal} = r/r_{\rm spal}$, $\rho_{\rm wind} = r/r_{\rm wind}$ et
$\rho_L = r/L$. The smallest of these three numbers indicates the
dominant effect.
Various $r_{\rm lim}$-probabilities are
shown in Tab.~\ref{big_table}.
\begin{table}[ht]
      \begin{tabular}{|ccc|ccc|}   \hline
         $L$(kpc) & $V_c$ (km~s$^{-1}$) & $\sigma_{\rm inel}$ (mb) &
         $R_{50\%}
         $(kpc)
         & $R_{90\%}
         $(kpc) & $R_{99\%}
         $(kpc) \\
         \hline
         $L_{10} \times 10$  & 0 & 0 & 6.3 $\times L_{10}$ & 19
$\times L_{10}$ &
         34 $\times L_{10}$\\
         $\infty$ & $10 \times V_{10}$ & 0 & $4.9/V_{10}$ & $18.6/V_{10}$ &
         $41/V_{10}$\\
         $\infty$ & 0 & $100\times \sigma_{100}$ & $3.7 /
         \sigma_{100}$ & $30.8 / \sigma_{100}$ & $318/ \sigma_{100}$\\
         \hline
         $5$ & 0 & 0 & 3.1 & 9.5 & 17\\
         $5$ & 10 & 50 & 2.05 & 6.8 & 13.1\\
         \hline
      \end{tabular}
      \caption{This table indicates the radius of several 
      $f$-surfaces, for several
      values of $L$, $V_c$ and $\sigma_{\rm inel}$.
      We have introduced $L_{10}\equiv L/10$~kpc, $V_{10} \equiv
      V_c/10$~km~s$^{-1}$ and $\sigma_{100}\equiv \sigma/100$~mb.}
      \label{big_table}
\end{table}
For a radioactive species, the spallations and the Galactic wind have
a negligible effect on propagation as long as $l_{\rm rad}$ 
(see Sec.~\ref{partie_rad}) is smaller
than $L$, $r_{\rm spal}$ and $r_{\rm wind}$.

                  %---------------------%

\subsection{The number of disk-crossings in the general case}

\begin{figure}[ht]
      \centerline{\includegraphics*[width=\columnwidth]{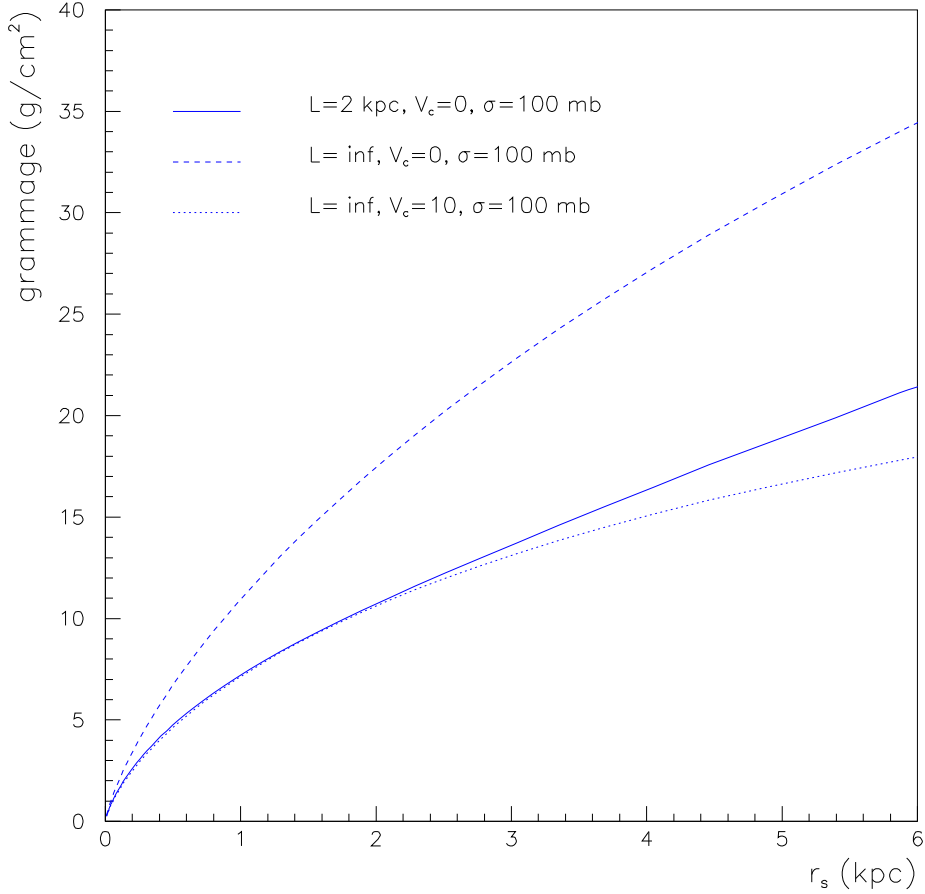}}
      \caption{Grammage crossed as a function of the origin, for some of the
      models discussed in the text and for a typical value of
      $K=0.03$~kpc$^2$~Myr$^{-1}$.}
      \label{fig:nb_cross}
\end{figure}

Several properties (energy losses, amount of reacceleration,
secondary-to-primary ratio) of the Cosmic Ray flux detected in the Solar
neighborhood are determined by the number of times a given Cosmic
Ray has crossed the disk since it was created.
The distribution of disk-crossings is computed in App.~C in the case of an
infinite diffusive volume and in the absence of Galactic wind.
In the most general situation, the mean number of crossings
(though not the entire distribution of crossing numbers)
can be computed as follows.
Each time a particle crosses the disk, it has a probability $p=2h
\sigma_{\rm inel} n_{\rm ISM}$ of being destroyed by a spallation.
The number $N(r)$ of surviving particles can thus be
obtained from $N_0(r)$, the number of particles diffusing without
spallations, as
\begin{displaymath}
       N(r) = N_0(r) \times \left( 1- p\right)^{n_{\rm cross}}\;,
\end{displaymath}
so that the number of crossing is readily obtained from the densities
with and without spallations as
\begin{equation}
       n_{\rm cross}(r) = \frac{\ln (N(r)/N_0(r))}
       {\ln( 1 - p)}\;.
       \label{number_of_crossings_2}
\end{equation}
Notice that this expression applied to
Eq.~(\ref{forme_utile}) leads to
Eq.~(\ref{final_nb_crossing}) when $L\rightarrow \infty$, $V_c = 0$, and
when $p$ is small.
As the surface density of the disk is $2hn_{\rm ISN} \sim 10^{-3} \unit{g}
\unit{cm}^{-2}$, the mean column density crossed by the particle
(called {\em grammage\/}) is given by
\begin{displaymath}
          \Sigma (r_s) = n_{\rm cross} (r_s)\times 2hn_{\rm ISM} \sim 
20 \unit{g}
          \unit{cm}^{-2}
         \times \frac{n_{\rm cross} (r_s)}{10^4}\;.
\end{displaymath}
The evolution of the grammage with the distance of the source is
displayed in Fig.~\ref{fig:nb_cross}. The effect of escape,
spallations and Galactic wind is shown.

As a cross-check, it can be noticed that in this approach, the mean grammage
\[
\langle \Sigma \rangle_{\rm spatial}=\int \Sigma (r_s)
\frac{d{\cal P} \left\{ r_s |  0  \right\}}{d^2\vec{r}_s} d^2\vec{r}_s
\]
yields the right order of
magnitude for the usual grammage derived from leaky box analysis
($\sim 9$~g~cm$^{-2}$).
Moreover, the knowledge of $n_{\rm cross} (r_s)$ allows to estimate
the magnitude of energy losses and reacceleration rates.

                 %---------------------%

\subsection{The energy dependence}
\label{subsec:energy}

The diffusion coefficient actually depends on energy.
A commonly used form (see \cite{Maurin02d} for a discussion) is
\begin{displaymath}
       K = K_0 \beta \left( \frac{\cal R}{\mbox{1 GV}} \right)^\delta
\end{displaymath}
where ${\cal R}$ stands for the rigidity, $K_0 \sim 0.01 - 0.1
\unit{kpc}^2 \unit{Myr}^{-1}$ and $\delta \sim 0.3-1$.
The previous results were given for $K = 0.03 \unit{kpc}^2 \unit{Myr}^{-1}$,
typical for a proton with an energy of 1 GeV.
This implies that the parameters  
$r_{\rm wind}$, $r_{\rm spal}$ are larger at higher energy.
They eventually become larger than $L$, so that at high energy escape 
dominates.
At low energy, the relative importance of spallation and convection 
can be evaluated by comparing $r_{\rm wind}$ and $r_{\rm spal}$.
However, it must be noticed that even when $r_{\rm wind}$ is greater than $r_{\rm spal}$,
the Galactic wind may have a non negligible effect on
the Cosmic Ray spatial origin because the cutoff due to $r_{\rm wind}$ is much
sharper (see Fig.~\ref{fig:influence_tot}). Moreover, the influence
of the Galactic wind on the spectra is important because of the
induced energy changes (adiabatic losses).

%%%%%%%%%%%%%%%%%%%%%%%%%%%%%%%%%%%%%%%%%%%%%%%%%%%%%%%%%%%%%%%%%%%%%%%%%%%%%%%%%%%%
%%%%%%%%%%%%%%%%%%%%%%%%%%%%%%%%%%%%%%%%%%%%%%%%%%%%%%%%%%%%%%%%%%%%%%%%%%%%%%%%%%%%

\section{Realistic source distribution}
\label{resultats_de_base}

\begin{figure}[ht]
\centerline{
\includegraphics*[width=0.8\columnwidth]{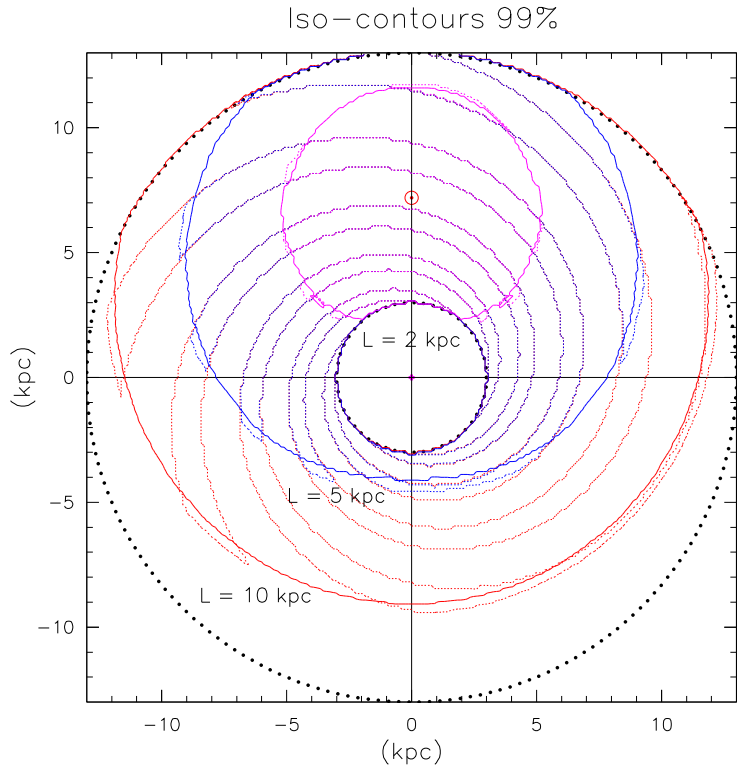}}
\caption{99\%-surfaces for $R=20$~kpc and
three cases, $L=2$~kpc, $L=5$~kpc and $L=10$~kpc.
}
\label{fig:dist_realiste}
\end{figure}

For the sake of definiteness, we will consider from now on that 
the cosmic ray sources for stable primaries are located in the disk and 
that their radial distribution 
$w(r_s)$ follows that of the pulsars and supernovae 
remnants, given by
\begin{equation}
      w_{\rm SN}(r_s)=\left(\frac{r_s}{R_\odot}\right)^{\alpha}
\exp\left(-\beta \times\frac{(r_s-R_\odot)}{R_\odot}\right)\;,
\label{case_batth}
\end{equation}
with $R_\odot = 8.5$~kpc,  $\alpha = 2$, $\beta = 3.53$ for \cite{case98}.
This distribution is now closer to the distribution adopted by
\cite{Strong98} ($\alpha = 0.5$ and $\beta = 1$), a flatter distribution
designed to reproduced radial $\gamma$-ray observations
(see Fig.~\ref{fig:dist_gamma}).
This distribution can be inserted in Eq.~\ref{proba_integree}, which 
is then used to compute the $f$-surfaces.
These surface are displayed in Fig.~\ref{fig:dist_realiste} for three cases 
($L=2$~kpc, $L=5$~kpc and $L=10$~kpc).
For large halos, the source distribution acts as a cutoff and
greatly limits the contributions from peripheric Galactic sources.

The results are not much affected by taking
an angular dependence into account.
Considering for example the spiral arms modelling of 
\citet{vallee02}, Fig.~\ref{fig:dist_realiste} shows that
the extension of the $f$-surfaces is almost not affected by
these small scale structures. In the rest of this paper,
the purely radial distribution (\ref{case_batth}) is assumed.

%%%%%%%%%%%%%%%%%%%%%%%%%%%%%%%%%%%%%%%%%%%%%%%%%%%%%%%%%%%%%%%%%%%%%%%%%%%%%%%%%%%%
%%%%%%%%%%%%%%%%%%%%%%%%%%%%%%%%%%%%%%%%%%%%%%%%%%%%%%%%%%%%%%%%%%%%%%%%%%%%%%%%%%%%

\section{Application to the propagation parameters deduced from the observed B/C ratio}

\label{BC_induced}
The previous sections present a complete description of the origin
of Cosmic Rays in a stationary diffusion model (energy losses and
gains are discarded). To each process by which a Cosmic Ray
may disappear before it reaches the Solar neighborhood is associated
a parameter: $L$ (escape through the top and bottom boundaries),
$r_{\rm wind}$ (convection), $r_{\rm spal}$ (destructive spallation).
The relative importance of these parameters may be measured by the
two quantities $\chi_{\rm wind} \equiv L/r_{\rm wind}$
and $\chi_{\rm spal} \equiv L/r_{\rm spal}$.
One can distinguish three regimes which determine the diffusion properties
of the system: i) the escape through the boundaries dominates
for $\chi_{\rm wind}\ll1$ and $\chi_{\rm spal}\ll1$; ii)
convection dominates for $\chi_{\rm wind}\gg1$ and
$\chi_{\rm wind}\gtrsim \chi_{\rm spal}$; iii) spallations dominate
for $\chi_{\rm spal}\gg1$ and  $\chi_{\rm wind}\ll\chi_{\rm spal}$.

We now use the sets of diffusion parameters consistent with the B/C data
given in \cite{Maurin02b} (hereafter MTD02) to evaluate realistic values 
for these quantities.

%%%%%############################%%%

\subsection{Evolution of $\chi_{\rm wind}$ and $\chi_{\rm spal}$ with $\delta$}
\label{chi_wind_chi_spal}
\begin{figure}[ht]
      \centerline{\includegraphics*[width=.75\columnwidth]{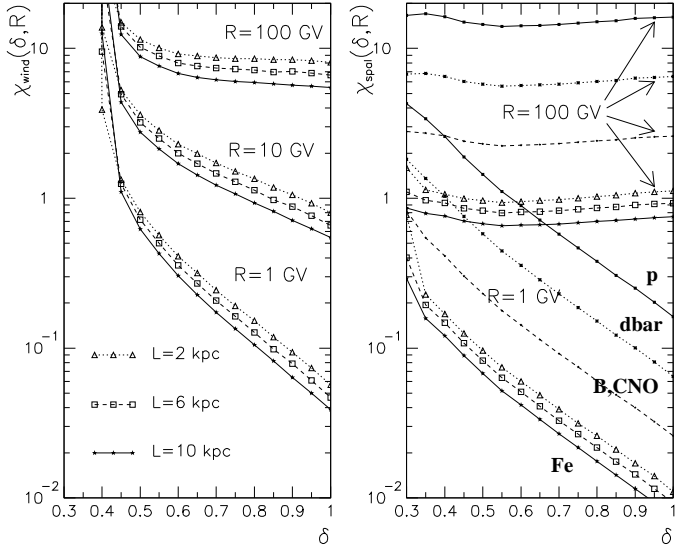}}
      \caption{Left panel: $\chi_{\rm wind}(\delta,{\cal R})$ as a 
function of the
      diffusion spectral index $\delta$ for different rigidities ${\cal R}$;
      from top to bottom, ${\cal R}=100$~GV, ${\cal R}=10$~GV and
${\cal R}=1$~GV.
      The parameter $\chi_{\rm wind}$, as well as $\chi_{\rm spal}$, is not very
      sensitive to the halo size $L$. Right panel: $\chi_{\rm
spal}(\delta,{\cal R})$
      as a function of $\delta$ for  ${\cal R}=100$~GV (upper curves) and
      ${\cal R}=1$~GV (lower curves) for four species: p ($\sigma\sim 40$~mb),
      $\bar{d}$ ($\sigma\sim 100$~mb), B-CNO ($\sigma\sim 250$~mb) and
      Fe ($\sigma\sim 700$~mb). For the latter species we plotted the same three
      $L$ values as in left panel. The behavior for other species
      is similar so that we only plotted the case $L=6$~kpc.}
      \label{fig:rw_rs_delta}
\end{figure}

In MTD02, we provide for each configuration $\alpha$
(source spectral index), $\delta$ (diffusion spectral index) and $L$
(diffusive halo size) the corresponding $K_0$, $V_c$ and $V_a$
(Alfv\'enic wind responsible for reacceleration) that fit best the ratio B/C.
In this study, $V_a$ is not very important since it only changes the energy
of the particles: a Cosmic Ray emitted at 1 GeV/nuc and gaining a few hundreds
of MeV/nuc during propagation will be detected at a slightly larger
energy, for which the results given here will not be very different.
This becomes even more true beyond a few GeV. 
Reacceleration will be ignored throughout
this study, as well as energy losses, for the same reason.
Moreover, the values of  $K_0$, $V_c$ and $V_a$ do not 
depend much on $\alpha$ (see Fig.~9 of MTD02), so that $\chi_{\rm wind}$ 
and $\chi_{\rm spal}$ depend mainly on $\delta$ and $L$.
They depend on rigidity, through $K(E)$, as
can be seen in Fig.~\ref{fig:rw_rs_delta} where
$\chi_{\rm wind}$ and $\chi_{\rm spal}$
are displayed as a function of $\delta$ for several species, several
values of $L$ and several rigidities.

The left panel displays $\chi_{\rm wind}(\delta,L)$ for
three rigidities: 1~GV, 10~GV and 100~GV.
Up to several tens of GV, convection is in competition with escape;
afterwards escape dominates.
The noticeable fact is that models corresponding to $\delta\lesssim 0.45$
are escape-dominated, whereas convection dominates
only for large $\delta$ at low energy.
It appears that all other parameters being constant, $\chi_{\rm
wind}$ is fairly independent of $L$.
(indicating a similar relative importance of
convection and escape for the models reproducing the B/C ratio, see MTD02).
However, the spatial origin does depend on $L$ and $r_{\rm wind}$ and not
only on their ratio.

The right panel of Fig.~\ref{fig:rw_rs_delta} plots $\chi_{\rm spal}(\delta,L)$
for ${\cal R}=100$~GV and 1~GV for various nuclei. Protons are the most abundant
species in Cosmic Rays.
Boron and CNO family are
important because they allow to constrain the value of the 
propagation parameters, e.g. through the B/C ratio. 
Last, the Fe group provides another test of the
secondary production {\em via\/} the sub-Fe/Fe ratio.
The evolution of $\chi_{\rm spal}$ for these species is conform to what
is very well known from earlier leaky box inspired studies: for heavier nuclei,
spallation dominates over escape and for this reason, the induced secondary
production is particularly sensitive to the low end of the grammage distribution.

To summarize, the left
panel shows the evolution from convection-domination to
escape-domination as a function of ${\cal R}$ and $\delta$, 
the effect of the wind being  negligible
above $\sim 100$~GeV  whatever $\delta$ ($\chi_{\rm wind}\gtrsim 10$).
The right panel gives the
evolution from spallation-domination to escape-domination as a function
of ${\cal R}$, $\delta$ and the species under consideration.
The effect of spallation is more important for heavy than for light 
nuclei, but this difference is too small to produce an evolution of
the average logarithmic mass for high energy ($\sim$~TeV) Cosmic Rays
\citep{Maurin02a}.

    %%%%%############################%%%
\subsection{Spatial origin in realistic diffusion models at 1 GeV/nuc}

From the previous discussion, it appears that spallations and Galactic
wind play a role at low energy.
The results will be shown for the particular value 1~GeV/nuc
which is interesting for various astrophysical problems. 
First, once modulated, it corresponds to about  the very lowest energy 
at which experimental set-ups have measured Galactic Cosmic Rays. 
Second, the low energy domain is the most favorable window  to observe $\bar{p}$
(resp. $\bar{d}$) from exotic sources (see companion paper \citet{Maurin02c}), as the background
corresponding to secondaries $\bar{p}$ (resp. $\bar{d}$) is reduced.
Last, these energies correspond to that of the enduring problem
of the diffuse GeV $\gamma$-ray radial distribution.
This was first quoted by \citet{Stecker77} and further investigated by
\citet{Jones79} taking into account the effect of a Galactic wind.

\begin{figure}[ht]
    \centerline{
    \includegraphics*[bb=100 240 553 625,clip,
    width=0.5\columnwidth]{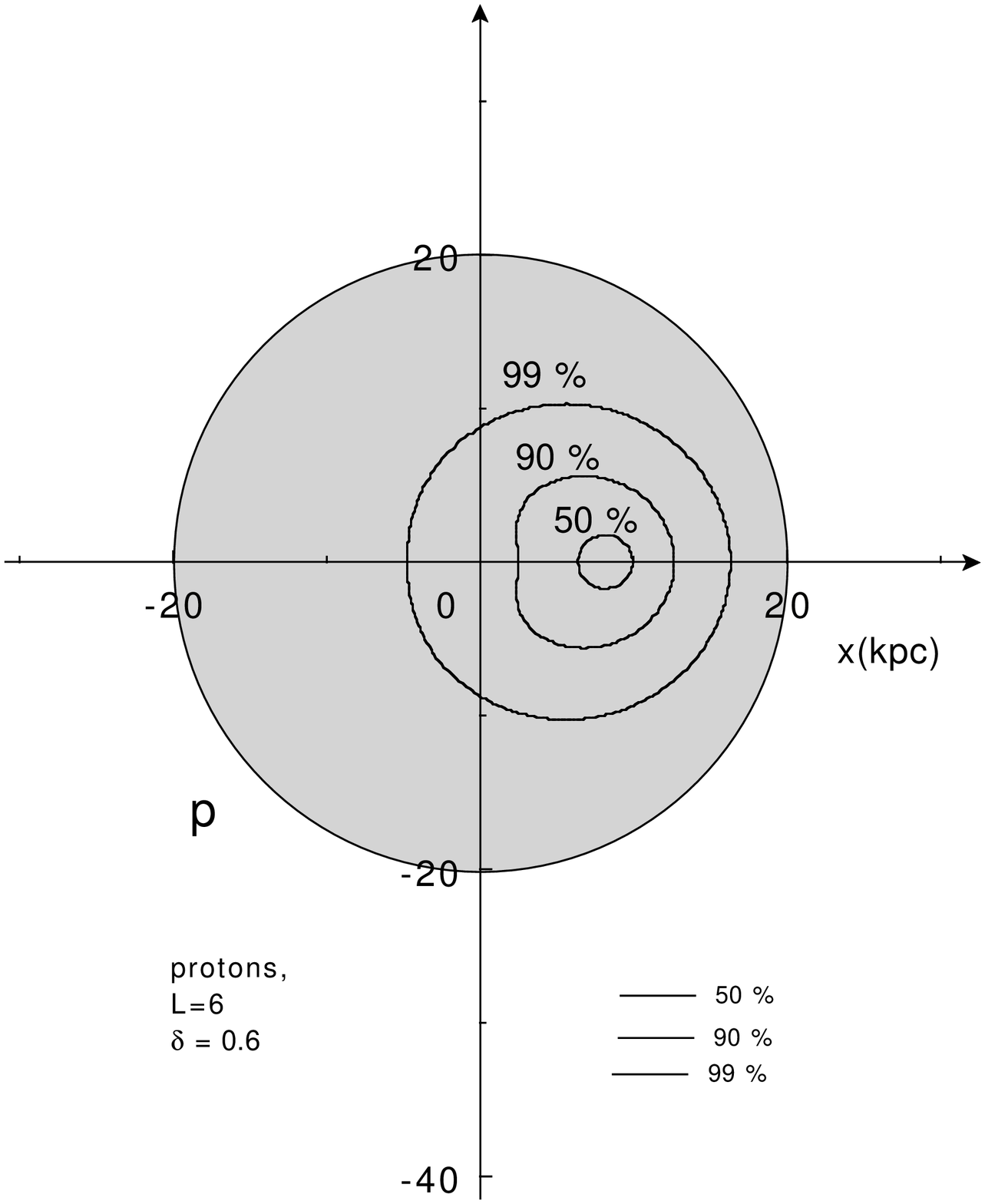}
    \includegraphics*[bb=100 240 553 625,clip,
    width=0.5\columnwidth]{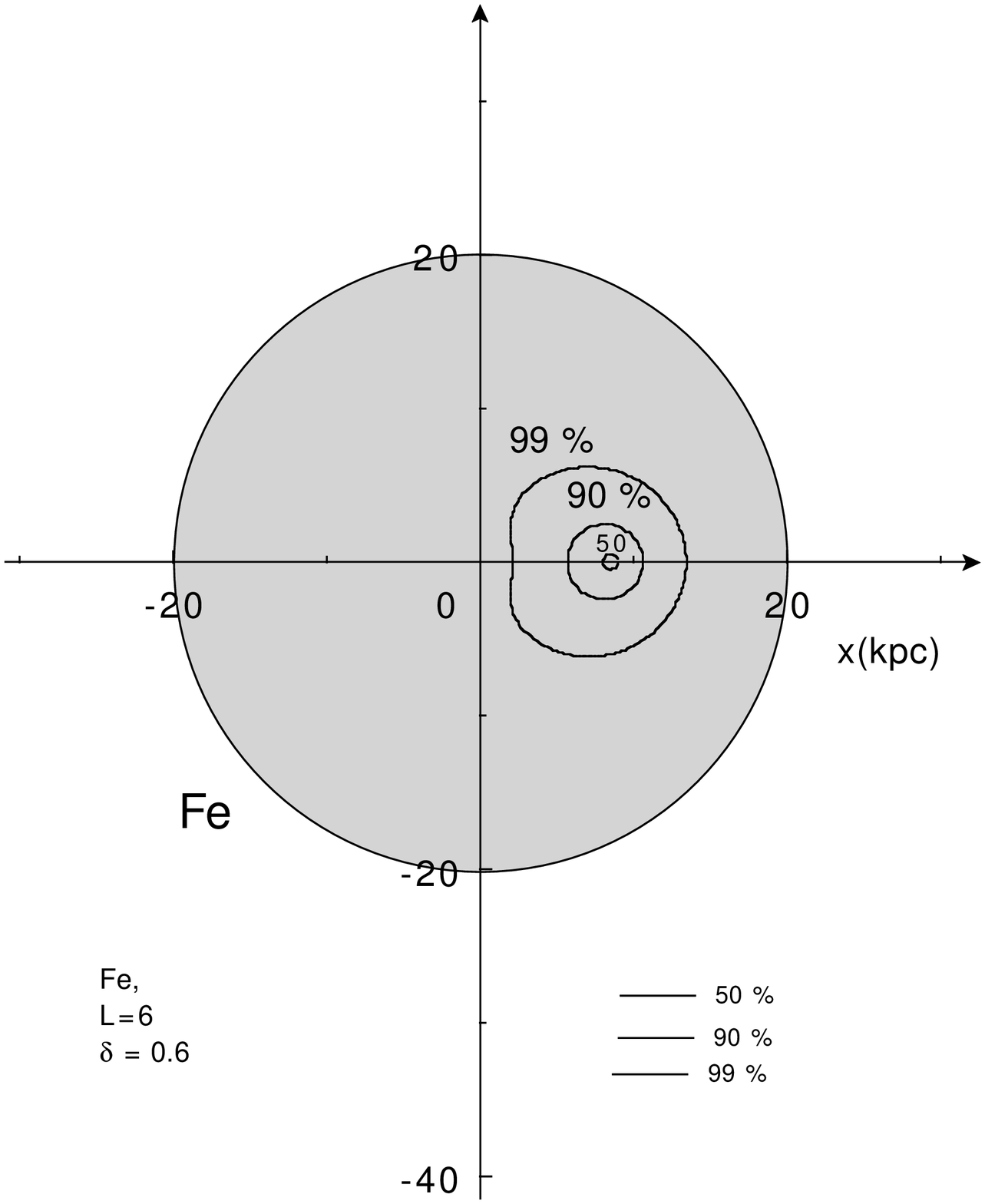}
    }
    \caption{(50-90-99)\%-surfaces (protons and
    Fe nuclei are considered), in a typical diffusion model with $L=6$~kpc and $\delta=0.6$.}
    \label{fig:dif_contours}
\end{figure}

\begin{figure}[ht]
\centerline{
\includegraphics*[bb=100 240 553 625,clip,
width=0.5\columnwidth]{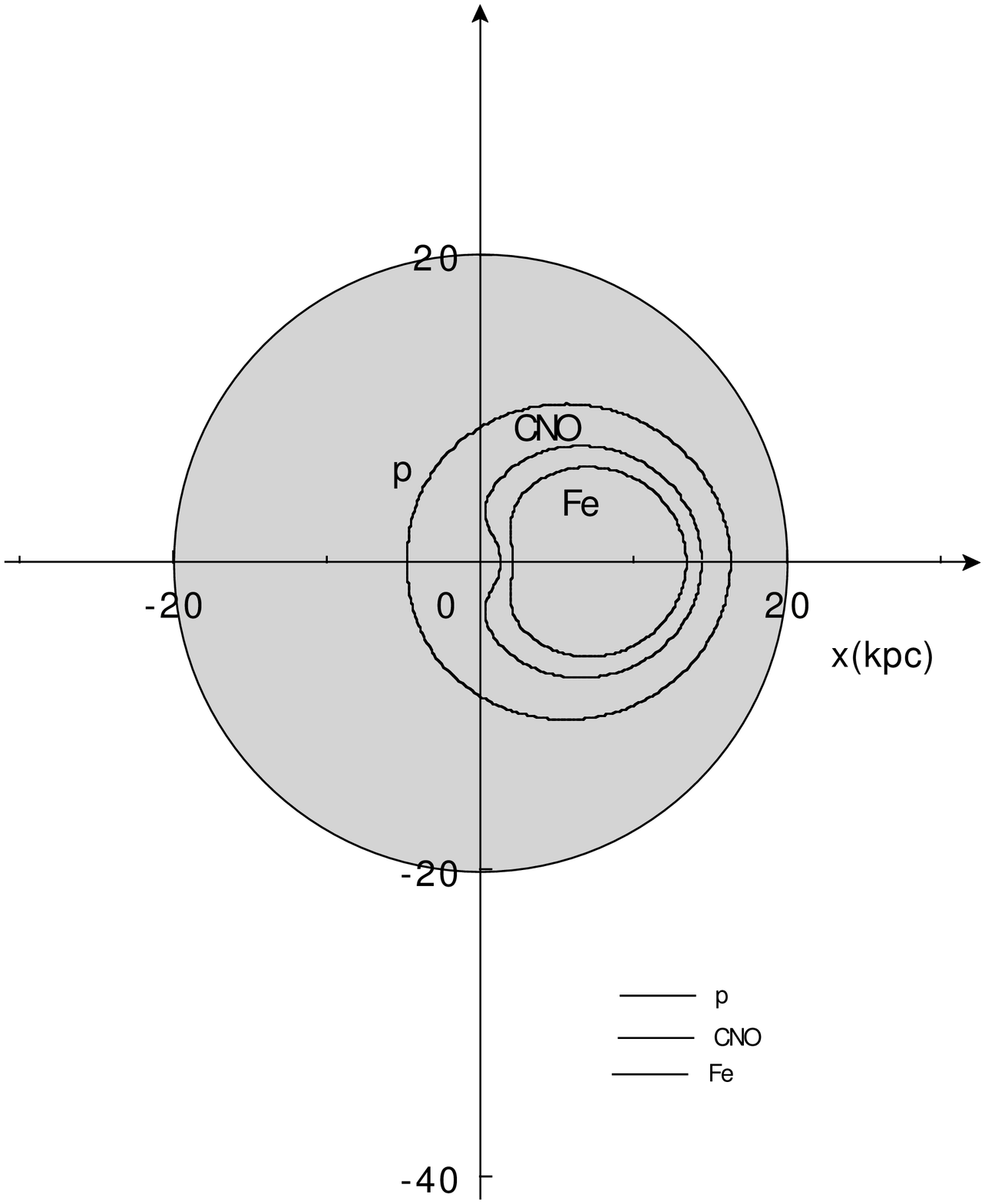}
\includegraphics*[bb=100 240 553 625,clip,
width=0.5\columnwidth]{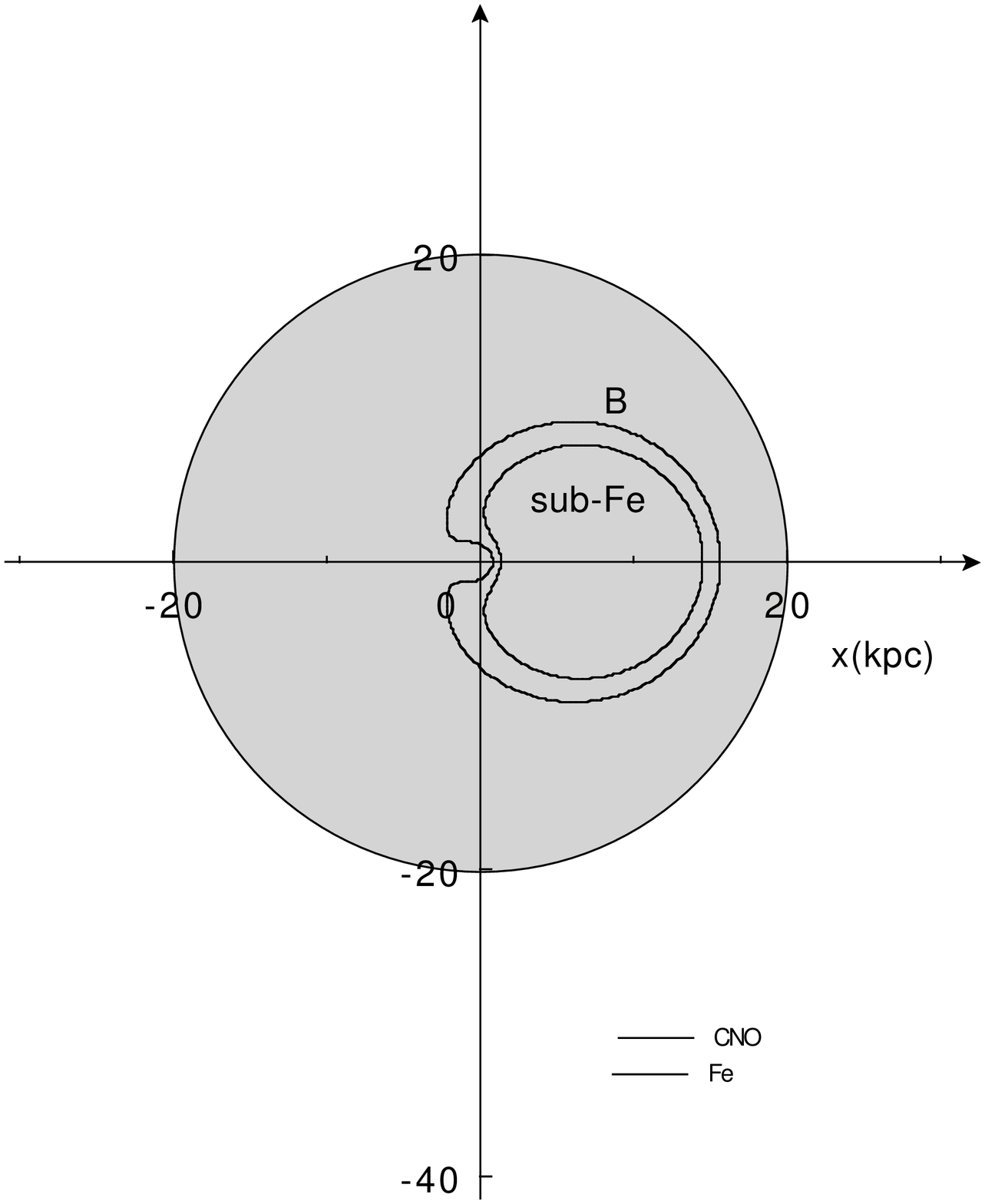}
}
\caption{99\%-surfaces for several species. The left panel corresponds to primary species
(protons, CNO and Fe) while the right panel corresponds to the 
progenitors of secondary species (B and sub-Fe), for $L=6$~kpc and $\delta=0.6$.}
\label{fig:effet_espece}
\end{figure}

\begin{figure}[ht]
\centerline{
\includegraphics*[bb=100 240 553 625,clip,
width=0.5\columnwidth]{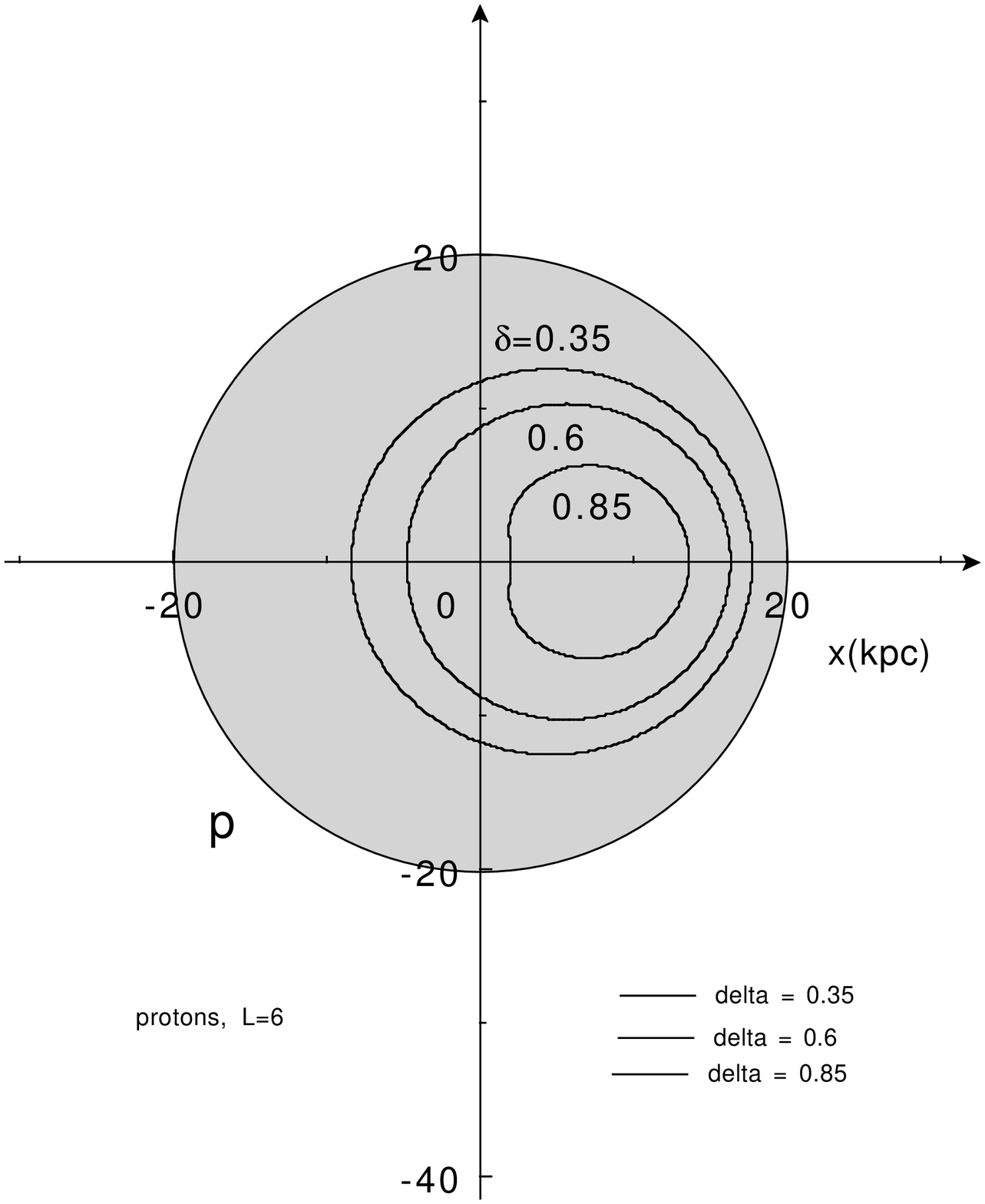}
\includegraphics*[bb=100 240 553 625,clip,
width=0.5\columnwidth]{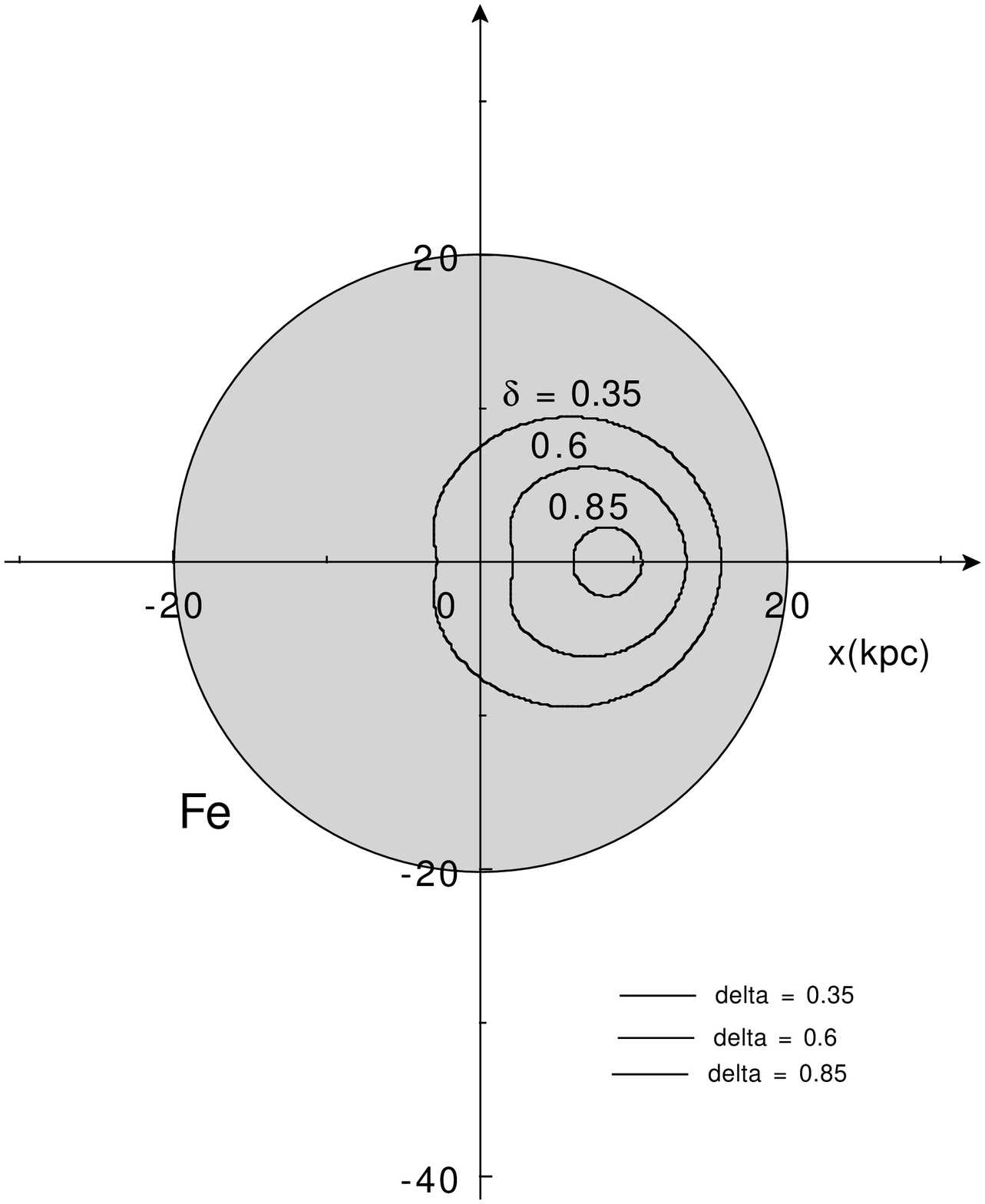}
}
\caption{99\%-surfaces for several $\delta$, in the case $L=6$~kpc.
The left panel corresponds to protons while the right panel corresponds
to Fe nuclei.}
\label{fig:effet_delta}
\end{figure}

\begin{figure}[ht]
\centerline{
\includegraphics*[bb=100 240 553 625,clip,
width=0.5\columnwidth]{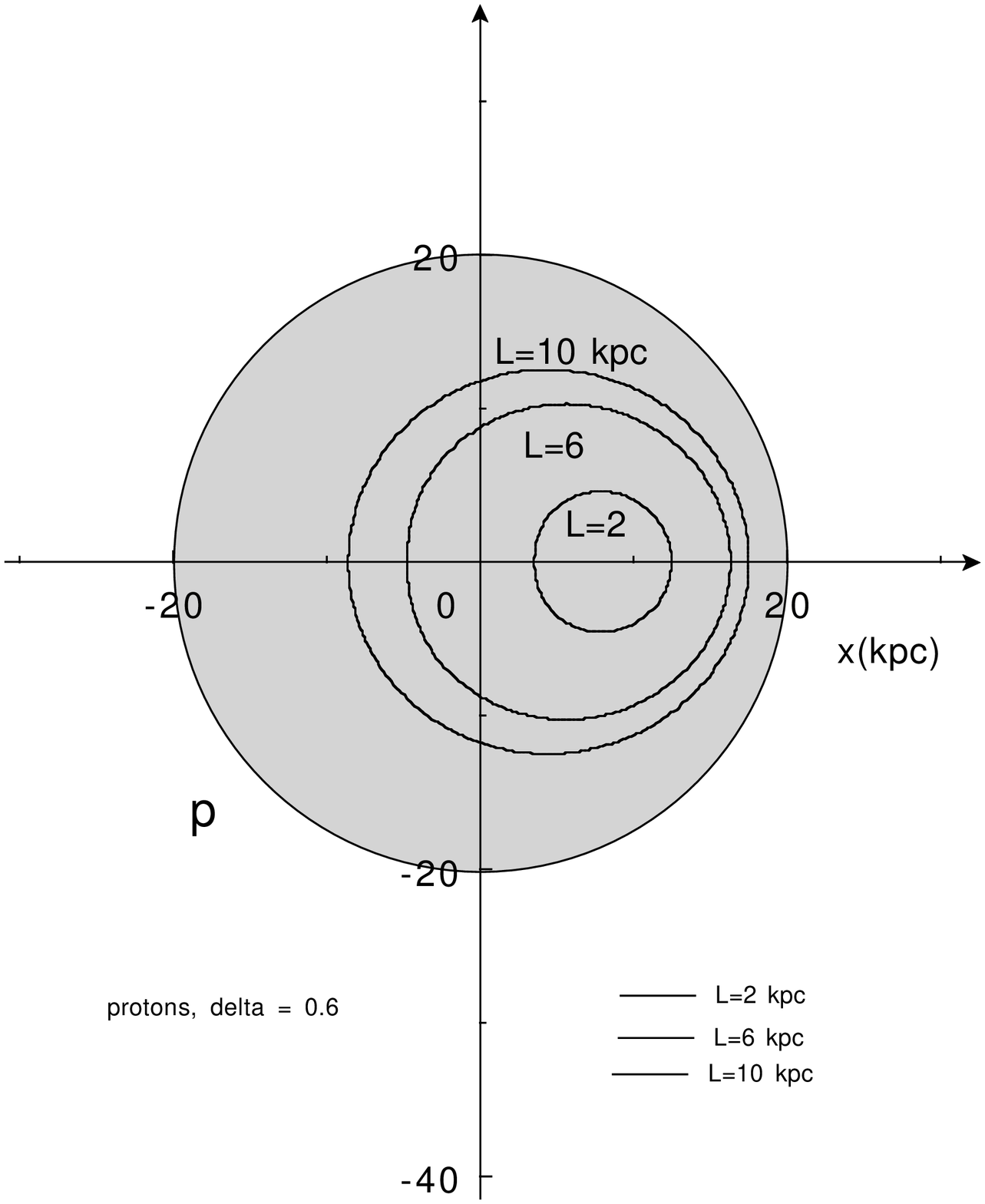}
\includegraphics*[bb=100 240 553 625,clip,
width=0.5\columnwidth]{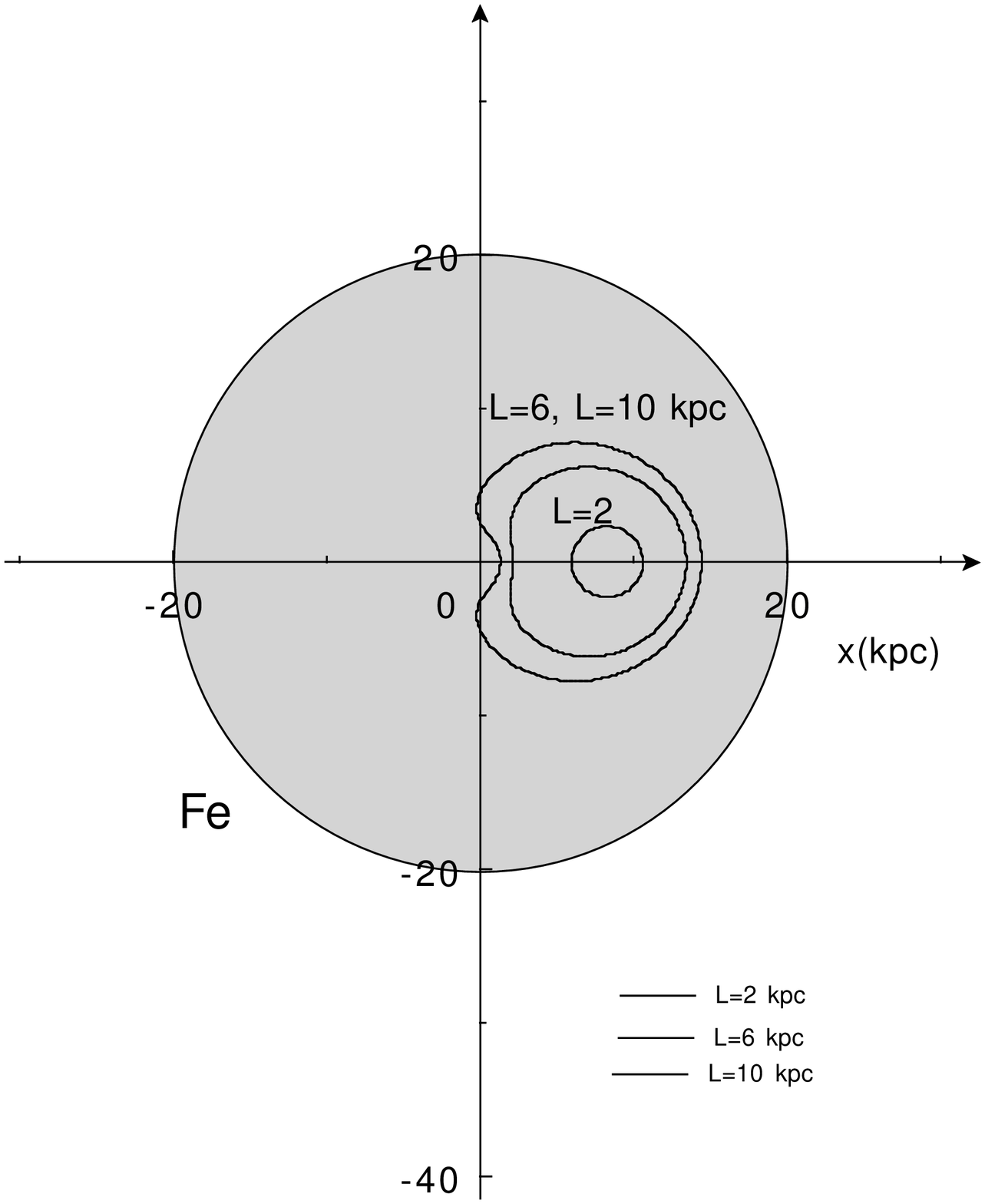}
}
\caption{99\%-surfaces for several $L$, in the case $\delta=0.6$.
The left panel corresponds to protons while the right
panel corresponds Fe nuclei.}
\label{fig:effet_L}
\end{figure}

From the sets of diffusion parameters that fit the B/C ratio,
the values of the parameters $r_{\rm wind}$ and $r_{\rm spal}$
are computed (see Tab.~\ref{last_table}) for
the four nuclei shown in Fig.~\ref{fig:rw_rs_delta} and for three
values of $\delta=[0.35,0.6,0.85]$.
%%%%%%%%%%%%%%%%%%%%%%%%%%%%%%%%%%%%%%%%%%%%%%%%%%%%%%%%%%%%%%%
\begin{table}[ht]
     \begin{center}
	\begin{tabular}{|cc||c|c|c|}   \hline
	    && $p$, $\bar{p}$ & B-CNO & Sub-Fe, Fe \\
	    && $\delta=0.35/0.6/0.85$& - & -\\\hline\hline
	    $L=10$~kpc &  $r_{\rm wind}$ (kpc) & $\infty$/5.17/1.6
	    &$\infty$/3.41/0.89 &$\infty$/3.26/0.83\\
	    & $r_{\rm spal}$ (kpc)& 33.5/10.2/4.0  & 4.21/1.07/0.35
	    &1.46/0.37/0.12\\
	    & $\sqrt{\langle r^2 \rangle}$(kpc)& 6.43/4.67/2.50 &
	    5.23/2.92/1.11 & 4.00/2.03/0.57\\
	    \hline
	    $L=6$~kpc &    $r_{\rm wind}$ (kpc) & $\infty$/3.64/1.15
	    &$\infty$/2.40/0.64 &$\infty$/2.30/0.60\\
	    &$r_{\rm spal}$ (kpc)& 24.7/7.4/2.9  & 3.10/0.80/0.26
	    &1.08/0.27/0.09\\
	    & $\sqrt{\langle r^2 \rangle}$ (kpc)& 4.93/3.5/1.88&
	    3.96/2.18/0.82 & 2.99/1.63/0.54\\
	    \hline
	    $L=2$~kpc & $r_{\rm wind}$ (kpc) & $\infty$/1.40/0.46
	    &$\infty$/0.92/0.26 &$\infty$/0.88/0.24\\
	    &$r_{\rm spal}$ (kpc)& 9.7/2.9/1.2  &1.21/0.31/0.10
	    &0.42/0.10/0.03 \\
	    & $\sqrt{\langle r^2 \rangle}$(kpc)& 2.07/1.44/0.78 &
	    1.63/0.87/0.33 & 1.21/0.57/0.19\\
	    \hline\hline
	\end{tabular}
	\caption{Values of $r_{\rm wind}$ and $r_{\rm spal}$ for the sets
	of parameters that, for a given $\delta$, give the best fit to the
	observed B/C ratio. The mean square value $\sqrt{\langle r^2 \rangle}$
	of the distance to the sources is also shown. 
	}
	\label{last_table}
     \end{center}
\end{table}
%%%%%%%%%%%%%%%%%%%%%%%%%%%%%%%%%%%%%%%%%%%%%%%%%%%%%%%%%%%%
From these values, the 50-90-99\%-surfaces are derived and displayed 
in Fig.~\ref{fig:dif_contours}, for protons and Fe nuclei.  
The effect  of $\delta$ (Fig.~\ref{fig:effet_delta}), of $L$ 
(Fig.~\ref{fig:effet_L}) and of the species (Fig.~\ref{fig:effet_espece}), 
are considered separately.
As regards the first two effects, the $f$-surfaces are smaller:
(i) for greater values of $\delta$, mainly 
because the effect of the wind is then greater, and
(ii) for small values of $L$, as in 
this case escape is more important.

As regards the last effect, it can first 
be seen from Fig.~\ref{fig:effet_L} that the heavier species come from a 
shorter distance (because the spallations are more important).
The secondary species can be treated simply by using a source function 
obtained by multuplying the primary density by the gas density. 
It would be straightforward to apply the previous techniques to a realistic 
gas distribution (taking into account, in addition to the fairly flat HI 
distribution, that of molecular H$_2$ and ionized HII which are more strongly peaked in the 
inner parts, see e.g. \citet{Strong98} for a summary and references)
and to infer the contours inside which the
secondaries are created. The corresponding $f$-surfaces are not shown here, as they 
would be quite similar to those of the primaries (see left panel).
What we do display in the right panel are the $f$-surfaces of the 
primaries that lead to given secondaries, as these progenitors 
determine the secondary-to-primary ratios (see Sec.~\ref{progenitors}).


                           %---------------------%

\subsection{Effective number of sources}

  From the previous results, it appears that only a fraction of the
sources present in the disk actually contribute to the flux in the
Solar neighborhood. In this paragraph we present the fraction $f_s$ of the
sources which are located inside given $f$-surfaces.
\begin{figure}[ht]
    \centerline{
    \includegraphics[width=0.85\columnwidth]{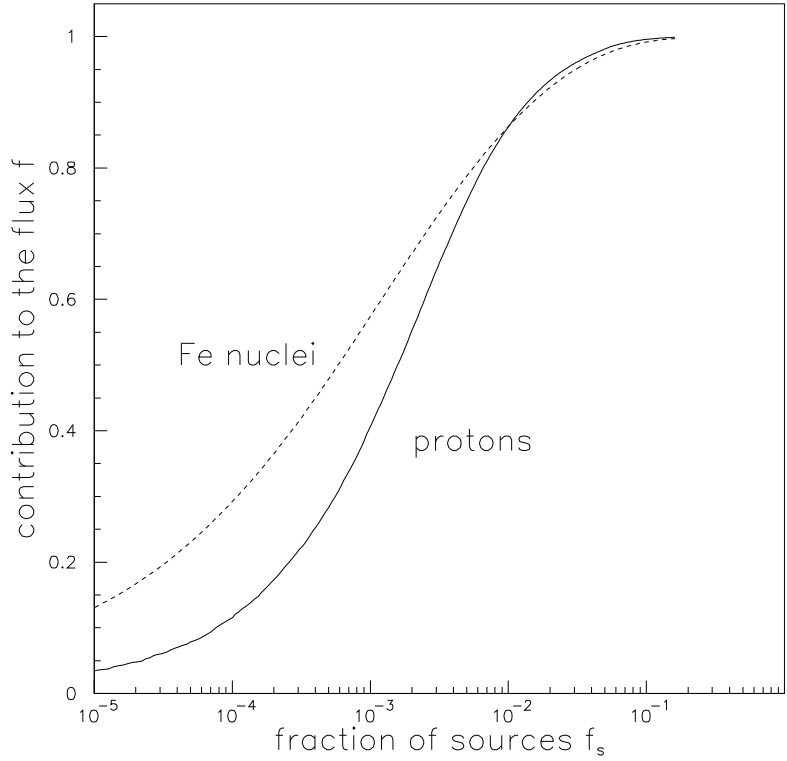}
    }
    \caption{Fraction $f_s$ (in \%) of the Galactic sources contributing to
    the  fraction $f$ of the Cosmic Ray flux at the Solar
    position, for protons and Fe nuclei, for the particular diffusion
    model $L=6$~kpc, $\delta = 0.6$.}
    \label{fig:dist_proba}
\end{figure}
This fraction is presented in Fig.~\ref{fig:dist_proba} for
the particular model $L=6$~kpc, $\delta = 0.6$, for protons and Fe
nuclei. It can be read that for example, it takes 7.6\% (resp. 1.5\%) of the
Galactic sources to make 90\% of the protons (resp. Fe nuclei) reaching
the Solar neighborhood.
%%%%%%%%%%%%%%%%%%%%%%%%%%%%%%%%%%%%%%%%%%%%%%%%%%%%%%%%%%%%%%%
\begin{table}[ht]
      \begin{center}
         \begin{tabular}{|cc||c|c|}   \hline
             && $p$, $\bar{p}$ & Fe \\
             && $\delta=0.35/0.6/0.85$ & -\\\hline\hline
             $L=10$~kpc &  90\% & 22.9/13.3/3.9  &9.6/2.2/0.24\\
                                 &  99\% & 56.7/40.7/16.4 &35/14.5/1.77\\\hline
             $L=6$~kpc  &  90\% & 14.3/7.6/2.3 & 5.63/1.5/0.17\\
                                &  99\%  & 40.9/27/9.6  & 22.2/9.2/1.48\\\hline
             $L=2$~kpc & 90\% & 2.9/1.6/0.83 &1.2/0.24/0.025\\
                                &99\% & 9.3/5.9/3.8  &4.7/1.77/0.27 \\
             \hline\hline
         \end{tabular}
         \caption{Fraction $f_s$ (in \%) of the Galactic sources 
	 contributing to a
         given fraction (90\% and 99\%) of the Cosmic Ray flux at the Solar
         position, for protons and Fe nuclei, for the diffusion models
         studied before.}
         \label{tab:dist_origin}
      \end{center}
\end{table}

Figure~\ref{fig:dist_proba} also shows that a very small fraction of
sources may contribute to a non negligible fraction of the fluxes.
For example, the sphere of radius  $r\sim 100$~pc centered on the
Solar neighborhood contains only $2.5 \, 10^{-5}$ of the sources 
but for $L=6$~kpc and $\delta=0.6$, it is responsible for about 5\%
of the proton flux and 18\% of the Fe flux.
The mean age of the Cosmic Rays is given by $\langle t \rangle \sim \langle
r^2\rangle/2K \sim$ 7-400~Myr (see Tab.~\ref{mean_age}).
For a supernova rate of three per century, the total  number of sources
responsible for the flux is $\sim 2 \, 10^5 - 10^7$.
This tells us that in models with the largest $\delta$, 18 \% of the Fe flux can be due to 
only 5 sources.
The approximation of stationarity and of continuous source
distribution is likely to break down with such a small number of 
sources. Conversely, for small values of $\delta$ 
(preferred by many authors), this approximation is probably better.
\begin{table}[ht]
     \begin{center}
	\begin{tabular}{|c|cc|}   \hline
	    & $p$, $\bar{p}$ &  Sub-Fe, Fe \\
	    & $\delta=0.35/0.6/0.85$&  -\\\hline\hline
	    $L=10$~kpc &  196 / 392 / 332 & 76 / 74 / 17\\
	    \hline
	    $L=2$~kpc  &  72/ 130 / 108 & 24/ 22 / 7\\
	    \hline\hline
	\end{tabular}
	\caption{Mean Cosmic Ray age $\langle t \rangle \sim \langle r^2 
	\rangle/2K$ in Myr for the models studied in this paper.}
	\label{mean_age}
     \end{center}
\end{table}
%%%%%%%%%%%%%%%%%%%%%%%%%%%%%%%%%%%%%%%%%%%%%%%%%%%%%%%%%%%%

\subsection{Radioactive species, $e^+$ and $e^-$ and the local bubble}

\citet{Donato02} emphasized that the existence
of a local underdensity ($n\lesssim 0.005$~cm$^{-3}$) around the Solar
neighborhood greatly affects the
interpretation of the flux of radioactive species at low energy
(we refer the interested reader to this paper for a deeper
discussion and references on the local interstellar medium).
The most important effect of this hole is that it exponentially
decreases the flux by a factor $\exp(-a/l_{\rm rad})$ ($a\lesssim 65-250$~pc is the radius
of the
local underdense bubble and $l_{\rm rad}$ is given by Eq.~(\ref{f_rad})).
This can be
easily understood as there is almost no gas in this region, hence no
spallations, leading to no secondary production.
The local bubble is obviously not spherical, but this approximation is
sufficient at this level. This attenuation factor is straightforwardly
recovered starting from the probability density as given in
Sec.~\ref{partie_rad}, if correctly normalized to unity. To this end,
the sources (here spallation of primaries on the interstellar medium) are
considered to be uniformly distributed in the disk, except in the
empty region $r<a$. The probability density is zero in the hole whereas
outside, it is given by
\begin{equation}
     d{\cal P}^{\rm hole}_{\rm rad} =
     \exp\left(\frac{a}{l_{\rm rad}}\right)
     d{\cal P}_{\rm rad} =
     \frac{\exp(-(r_s-a)/l_{\rm rad})}{2\pi \,r_s .\, l_{\rm rad}}
     \,  d^2\vec{r}_s\;.
\end{equation}
The quantity ${\cal P}^{\rm hole}_{\rm rad} (r_s < r_{\rm lim}| O )$
is obtained directly from the no hole case (see Sec.~\ref{partie_rad})
by replacing $r_{\rm lim}$ by $r^a_{\rm lim}=r_{\rm lim}+a$.
It means that the sources that contribute to the fraction 
$f=(50-90-99)$\% of the
flux of the radioactive species are located between $a$ and
$r^a_{\rm lim}=(0.7-2.3-4.6)\times l_{\rm rad}+a$.
Hence, the hole only plays a marginal role for the origin of a 
radioactive species
(unless $l_{\rm rad}\lesssim a$), whereas the result for the
flux is dramatically different.

We saw in a previous section that the high energy $e^+$ and $e^-$
behave like unstable species. Their
typical length $r_{\rm loss}$ can be compared to $l_{\rm rad}$
\begin{equation}
    \left\{
    \begin{array}{l}
	\displaystyle
	\frac{l_{\rm rad} \mbox{(kpc)}}{\sqrt{\tau_0/1 \unit{Myr}}}\approx 0.12 \; \sqrt{\frac{\cal
	R}{1\unit{GV}}} \times
	\sqrt{\frac{K_0({\cal
	R}/1\unit{GV})^\delta}{0.03\unit{kpc}^2\unit{Myr}^{-1}}}
	\;\;;\vspace{0.2cm}\\\displaystyle
	r_{\rm loss} \mbox{(kpc)} \approx  \sqrt{\frac{1 \unit{GeV}}{E}} \times
	\sqrt{\frac{K_0({\cal R}/1\unit{GV})^\delta}{0.03 
	\unit{kpc}^2\unit{Myr}^{-1}}}\;\;.
    \end{array}
    \right.
\end{equation}
The dependence in the propagation model is similar for both expressions
and is
contained in the last term.
There is a big difference, though, as the  typical distances travelled
by radioactive nuclei scale as $\sqrt{\cal R}$, whereas they scale as
$1/\sqrt{E}$ for electrons and positrons.

\begin{figure}[ht]
    \centerline{\includegraphics*[width=0.8\columnwidth]{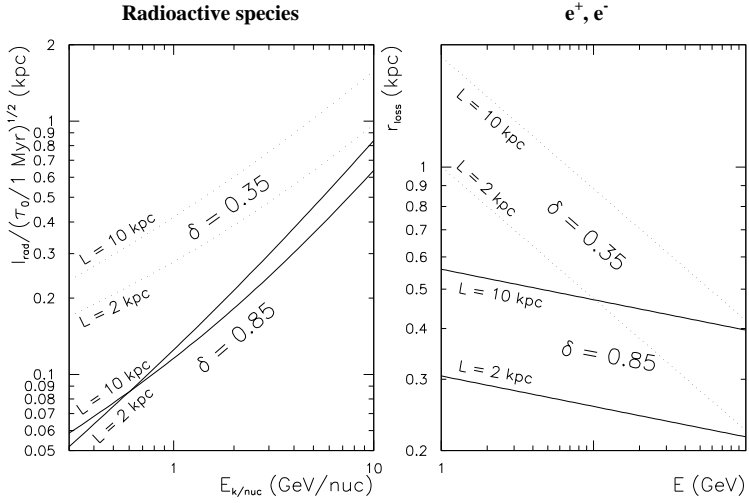}}
    \caption{Realistic values of $l_{\rm rad}/\sqrt{\tau_0/{\rm 
    1~Myr}}$ and
    $r_{\rm loss}$ for two extreme halo sizes $L$ and diffusion slope
    $\delta$. As all results in this section, propagation parameters
    fit B/C and are taken from MTD02.}
    \label{fig:rad_Realiste}
\end{figure}

Realistic values for $l_{\rm rad}$ and $r_{\rm loss}$ are presented
in Fig.~\ref{fig:rad_Realiste}.
At high energy, the Lorentz factor enhances the lifetime
of radioactive nuclei, making their origin less local, whereas the
energy losses are increased for electrons and positrons, making their
origin more local (99-90-50\% of 100 GeV $e^+e^-$ 
come from sources located in a thin disk with radius $r^{e^+e^-}_{\rm lim}
\approx 1.1-0.55-0.38$~kpc).
For 100 GeV $e^+$ and $e^-$, all models fitting B/C have the 
same value for $K_0{ \cal R}^\delta$, because at 100 GeV/nuc, spallations
and convection 
are negligible \citep{Maurin02b}. As a consequence, for GeV energies, 
$r_{\rm loss}$ increases more rapidly
for small $\delta$ than it does for larger $\delta$. To study the effect
of local contributions to the spectra of $e^+$ and $e^-$, \citet{Aharonian95}
used the value $\delta=0.6$ and compared to other works with $\delta=0$.
As these authors noticed, the modelling in the whole energy spectrum
is $\delta$ dependent, but our study gives the range compatible with B/C
studies.

Finally, radioactive nuclei are a very important tool for Cosmic Ray physics. 
They come from a few hundreds of parsec, and their fluxes are very 
sensitive to the presence of a local underdense bubble,
through the attenuation factor $\kappa \equiv \exp(-a/l_{\rm rad})$. 
For example, for a typical bubble of size
$a=100$~pc and an energy 800~MeV/nuc (interstellar energy),
$\kappa_{0.35}\approx\exp(-0.33 \sqrt{1 \unit{Myr}/\tau_0})$ if $\delta=0.35$,
whereas 
$\kappa_{0.85}\approx\exp(-\sqrt{1 \unit{Myr}/\tau_0})$. 
With $\tau_0^{\rm 10Be}=2.17$~Myr, $\tau_0^{\rm 26Al}=1.31$~Myr 
and $\tau_0^{\rm 36Cl}=0.443$~Myr, it leads to
$\kappa_{0.35}^{\rm Be,~Al,~Cl}\approx0.80-0.75-0.61$ and
$\kappa_{0.85}^{\rm Be,~Al,~Cl}\approx0.51-0.42-0.22$.
For $^{14}$C, the attenuation is $\kappa^{14\rm C}\ll 1$ around 1 
GeV/nuc, so that this species is heavily suppressed. 
However, it should be present around 10-100 GeV/nuc  
(as $\kappa^{14\rm C}\sim 1$ at these energies), with the advantage 
that solar modulation is less important at these energies.

The flux of radioactive species directly characterizes the 
local diffusion coefficient $K_0$ if the local environment is specified.
This would in turn allow to break the degeneracy in propagation parameters
that one can not avoid at present. Last, even though the surviving fraction of a
radioactive does depend on the halo size $L$, we emphasize that it is a
very indirect way to derive the propagation parameters. 
In the forthcoming years, new measurements of radioactive
species that do not depend on $L$ (e.g. by {\sc pamela} and {\sc ams}) 
should provide a promising path to update our vision of cosmic ray propagation.

%%%%%%%%%%%%%%%%%%%%%%%%%%%%%%%%%%%%%%%%%%%%%%%%%%%%%%%%%%%%%%%%%%%%%%%%%%%%%%%%%%%%
%%%%%%%%%%%%%%%%%%%%%%%%%%%%%%%%%%%%%%%%%%%%%%%%%%%%%%%%%%%%%%%%%%%%%%%%%%%%%%%%%%%%

\section{Summary, conclusions and perspectives}

\begin{figure}[ht]
      \centerline{
      \includegraphics[width=0.85\columnwidth]{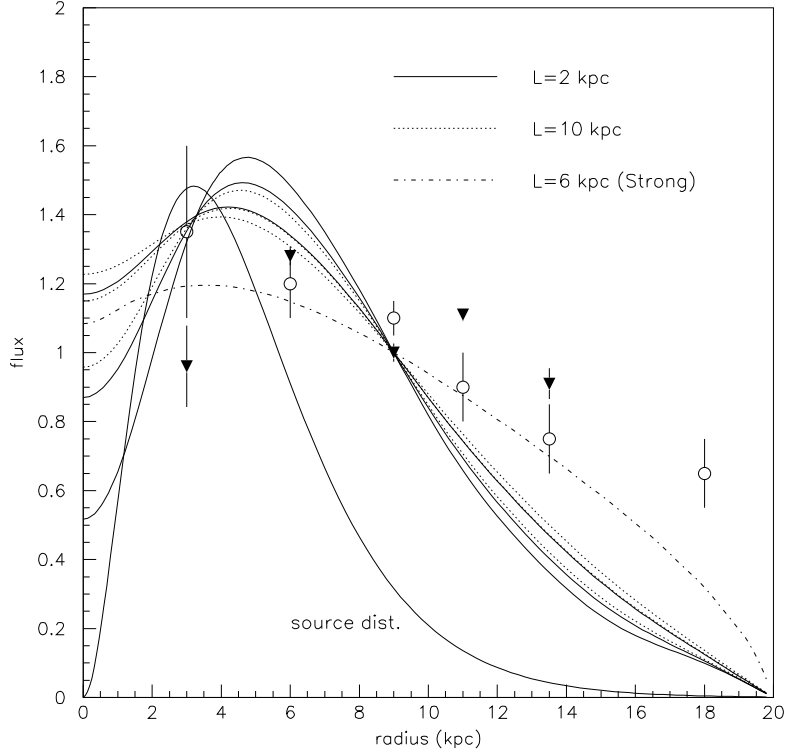}
      }
      \caption{Radial distribution of the proton flux for the models
      discussed in this study, compared to the source radial
      distribution of \citet{case98} given Eq.~(\ref{case_batth}).
      For each of the values $L=2$~kpc and $L=10$~kpc, the three values
      $\delta=0.35$, 0.6 and 0.85 are presented, the flatter
      distribution corresponding to the lower $\delta$.
      Also shown is the gamma-ray emissivity per gas atom 
      ({\sc cos-B} \citet{bloemen89}, 
      which is proportional to the proton flux, as given by
      {\sc cos-B} (open circles, \citet{bloemen89}) and {\sc egret}
      (triangles, \citet{strong_mattox96}),
      along with the proton flux obtained with the \citet{Strong98} distribution
      (see Sec.~\ref{resultats_de_base}).}
      \label{fig:dist_gamma}
\end{figure}

The question of the source distribution is very present
in Cosmic Ray physics. With the occurrence of the old problem of short
pathlengths distribution in leaky box models (see for example 
\citet{webber98}), \cite{lezniak79} studied a
diffusion model with no-near source in the Solar neighborhood.
Later, \cite{webber93a, webber93b} propagated $\delta$-like sources
with diffusion generated by a Monte Carlo random walk for the same
purpose. \cite{Brunetti00} follow this line but they introduce
random walks in a more realistic environment, i.e. circular,
elliptical and spiral magnetic field configurations. In a more formal
context, \cite{lee79} used a statistical treatment of means and
fluctuations (see references therein) to characterize amongst others
the possibility that nearby recent sources may dominate the flux of
primaries.
Finally, it is known that the present Cosmic Ray models are not able
to reproduce accurately for example proton-induced $\gamma$-rays measurements.
To illustrate this point, we plot in Fig.~\ref{fig:dist_gamma} the
radial distribution of protons
obtained with the same diffusion parameters as used above.
None of the models shown match the data.
One is left with two alternatives: either modifying the source 
distribution (for example, 
the distribution of \citet{Strong98} yields a better agreement),
or giving up the assumption that the diffusion
parameters apply to the Galaxy as a whole \citep{breit02}.
It is thus of importance to understand to what extent the Cosmic
Rays detected on Earth are representative of the distribution of the sources
in the whole Galaxy.

We provide an answer to this question under the two important hypotheses
that the source distribution is continuous and that we have reached a
stationary regime:
most of the Cosmic Rays that reach the Solar neighborhood were
emitted from sources located in a rather small region of the Galactic
disk, centered on our position.
The quantitative meaning of ``rather
small" depends on the species as well as on the values of the
diffusion parameters.
For the generic values $\delta = 
0.6$ and $L=6$ kpc chosen among the preferred values fitting B/C (see 
Sec.~\ref{BC_induced}), 
half of the protons come from sources nearer than 2 kpc, 
while half of the Fe nuclei come from sources nearer than 500 pc.
Another way to present this result is to say that the
fraction of the whole Galactic source distribution that actually
contributes to the Solar neighborhood Cosmic Ray flux can be rather 
small.
For the generic model just considered, 8 \% (resp. 1.5 \%) of the sources are required 
to account for 90 \% of the proton (resp. Fe) flux.
These fractions are smaller for higher $\delta$ and smaller $L$.
To summarize, 
the observed Cosmic Ray primary composition may be dominated by sources within a few kpc,
so that a particular care should be taken to model these source, 
spatially as well as temporally \citep{Maurin03b}.

Independently of all the results, this study
could be used as a check for more sophisticated Monte Carlo simulations
that will certainly be developed in the future to explore inhomogeneous
situations.
Several other consequences deserve attention.
First, the results may point towards the necessity to go beyond the
approximations of both continuity and stationarity. In particular, it
could be that only a dynamical model, with an accurate
spatio-temporal description of
the nearby sources, provides a correct framework to understand the  propagation
of Galactic Cosmic Rays. The contribution
from nearby sources would be very different in the low energy
($\sim$~GeV/nuc) or in the high energy regime ($\sim$~PeV) compared to
the stationary background.
Second, as discussed in Sec.~\ref{progenitors},
the diffusion parameters derived from the observed B/C ratio
have only a local validity, and one should be careful before applying
them to the whole Galaxy, since the Cosmic Rays are blind to most of it.

%%%%%%%%%%%%%%%%%%%%%%%%%%%%%%%%%%%%%%%%%%%%%%%%%%%%%%%%%%%%%%%%%%%%%%%%%%%%%%%%%%%%
%%%%%%%%%%%%%%%%%%%%%%%%%%%%%%%%%%%%%%%%%%%%%%%%%%%%%%%%%%%%%%%%%%%%%%%%%%%%%%%%%%%%
%%%%%%%%%%%%%%%%%%%%%%%%%%%%%%%%%%%%%%%%%%%%%%%%%%%%%%%%%%%%%%%%%%%%%%%%%%%%%%%%%%%%

\section*{Acknowledgments}
This work has benefited from the support of PICS 1076, CNRS and of
the PNC (Programme National de Cosmologie).
We warmly thank Eric Pilon for his expertise on asymptotic
developments. 
We also thank the anonymous referee for his pertinent suggestions.

%%%%%%%%%%%%%%%%%%%%%%%%%%%%%%%%%%%%%%%%%%%%%%%%%%%%%%%%%%%%%%%%%%%%%%%%%%%%%%%%%%%%
%%%%%%%%%%%%%%%%%%%%%%%%%%%%%%%%%%%%%%%%%%%%%%%%%%%%%%%%%%%%%%%%%%%%%%%%%%%%%%%%%%%%

\appendix

%%%%%%%%%%%%%%%%%%%%%%%%%%%%%%%%%%%%%%%%%%%%%%%%%%%%%%%%%%%%%%%%%%%%%%%%%%%%%%%%%%%%
\section{General solutions of the diffusion equation in cylindrical geometry}
\label{sol_cyl}
\subsection{General solution for our diffusion-convection model}

For a primary species, the differential density 
(in energy) $N(r,z)$ is a solution of the equation 
(see for example \cite{Maurin02d} and references
therein)
\begin{equation}
          \label{EQUATION GENERALE}
         {\cal L}_{\rm diff}N(r,z) +\Gamma_{\rm rad} N(r,z)
         +2h\delta(z)\; n_{\rm ISM}.\sigma_{\rm inel}.v \;N(r,z) =
          2h\delta(z)w(r) \;\;,
\end{equation}
with ${\cal L}_{\rm diff} =V_{c} \frac{\partial}{\partial z}
       -K\left(\frac{\partial^{2}}{\partial z^{2}}+
       \frac{1}{r}\frac{\partial}{\partial r}
       (r\frac{\partial}{\partial r})\right)$.
The various terms in Eq.~(\ref{EQUATION GENERALE}) correspond respectively
to (i) a differential operator ${\cal L}_{\rm diff}$ describing convection
$V_c$ out from the Galactic plane and isotropic diffusion $K$ throughout
the confinement volume; (ii) radioactive decay of the
unstable nucleus in the whole Galaxy;
(iii) destruction in flight $n_{\rm ISM}.\sigma_{\rm inel}.v$ when crossing
the gaseous thin disk of constant height $2h$ and constant density
$n_{\rm ISM}$; (iv) a source term in the thin disk.
The solution is determined by the boundary conditions of the problem
where Cosmic Rays freely escape, i.e. we demand $N(r=R,z)=N(r,|z|=L)=0$
($L$ is the half-height of the diffusive halo, $R$ the radial
extension of the Galaxy).
The solution in the disk ($z=0$) is given by
% \begin{equation}
%     \left\{
%     \begin{array}{l}
% 	\displaystyle 
% 	 N(r,0)=
%        \sum_{i=0}^{\infty}\,  {\frac{{\cal Q}_i}{A_i}} \,
%       ~ J_0
%       \left(\zeta_i\frac{r}{R} \right)
%        \label{sol_gen}\\
%        \displaystyle 
%        A_i= 2h \Gamma_{\rm inel} + V_c +
%       KS_i \, \coth \left(\frac{S_iL}{2} \right) \;\;\;\;\;\;\mbox{and}\;\;\;\;\;\; S_i^2=\frac{4 \zeta_i^2}{R^2} + \frac{V_c^2}{K^2}
%       + 4\frac{\Gamma_{\rm rad}}{K}~.
% 	  \end{array}
%     \right.
% \end{equation}
\begin{equation}
    N(r,0)= \sum_{i=0}^{\infty}\,  {\frac{{\cal Q}_i}{A_i}} \,
    ~ J_0 \left(\zeta_i\frac{r}{R} \right)
    \label{sol_gen}
\end{equation}
with
\begin{equation}
    A_i= 2h \Gamma_{\rm inel} + V_c +
    KS_i \, \coth \left(\frac{S_iL}{2} \right) 
    \;\;\;\mbox{and}\;\;\;
    S_i^2=\frac{4 \zeta_i^2}{R^2} + \frac{V_c^2}{K^2}
    + 4\frac{\Gamma_{\rm rad}}{K}~.
\end{equation}
and ${\cal Q}_i$ is the Bessel transform of the source distribution
(which may depend on the density of another species, in particular for
secondary species).

                           %---------------------%

For a primary point source, $w(r) = \delta(r)/(2\pi r)$
and we find in the disk ($z=0$)
\begin{equation}
       N_\delta^{\rm cyl}(r,0)=\sum_{i=1}^{\infty}
       \left\{
       \frac{1}{\pi J_1^2(\zeta_i)R^2A_i}
       \right\}\times J_0\left(\zeta_i\frac{r}{R}\right)\;.
       \label{sol_prim}
\end{equation}

                           %---------------------%

The generic solution for secondaries can be straightforwardly
derived from that of primaries (e.g. \cite{Maurin01}),
\begin{equation}
      \label{sec.}
      N^{\rm cyl}(r,0)=2h\Gamma_{{\rm prim}\rightarrow{\rm sec}}\times
      \sum_{i=1}^{\infty}\left\{
      \frac{1}{\pi J_1^2(\zeta_i)R^2 A^{\rm prim}_i A^{\rm sec}_i}
      \right\}J_0\left(\zeta_i\frac{r}{R}\right)\;.
\end{equation}
We use $\Gamma_{{\rm prim}\rightarrow{\rm sec}}=
n_{\rm ISM}.v.\sigma_{{\rm prim}\rightarrow{\rm sec}}$.
The distinction between $A^{\rm prim}_i$ and $A^{\rm sec}_i$ is necessary 
since both species have different destruction rates and rigidities.
%%%%%%%%%%%%%%%%%%%%%%%%%%%%%%%%%%%%%%%%%%%%%%%%%%%%%%%%%%%%%%%%%%%%%%%%%%%%%%%%%%%%

\section{Numerical evaluation of the point source solution in Bessel basis}
\label{numerique}
In practice, the infinite sums above are truncated to some order $n_{\rm tronc}$,
chosen as a compromise between accuracy (good
convergence of the series) and computer time.
In the case of a point source $\delta(\vec{r})$, the profiles are
singular near the
source and the convergence of the series appears to be very slow.
A few methods are presented to speed up this convergence.

\subsection{Softening of the source term}
First, the source term may be spread out on a radius $a$, by
replacing the $\delta$-function
by $w(r) = \theta(a-r)/(\pi a^2)$
for which an extra $2 J_1(\zeta_i a/R)/(\zeta_i a/R)$ term
appears in the Bessel transform.
With a judicious choice of the parameter $a$, the solution is very close 
to the original for $r\gg a$, but convergence is much faster due to the extra
$1/\zeta_i$ factor.

                           %*****%

\subsection{Sum representation: comparison to a known function}
\label{ansatz}

Part of the difficulty to evaluate numerically the Bessel expansions 
comes from the fact that the resulting functions are singular at the 
source position. If we know a reference function $f^{\rm ref}(r)$ which exhibits the same 
singularity and for which the Bessel coefficients 
$\mu_i^{\rm ref}$ are known, it is then judicious to write the density ($\rho \equiv 
r/R$) as
\begin{equation}
    N^{\rm cyl}_{\delta}(r,0)=
    \left[ \sum_{i=1}^{n_{\rm tronc}} \left(N^{\rm cyl}_i(0)-
    \mu_i^{\rm ref}\right)
    J_0(\zeta_i \rho) \right] +f^{\rm ref}(\rho)\,.
    \label{eq_ansatz}
\end{equation}
where the singularity is entirely contained in the $f$ term, so that 
the Bessel expansion has been regularized.
The choice of $f$ may be guided by the  behaviors observed in Sec.~\ref{bound}. 
For $L\sim R$, the solutions should be quite similar to the 
solution in spherical geometry, given by $f^{\rm ref}(r)=(1-\rho)/(4\pi \rho K R)$.
For very small halo size ($L< 1$~kpc), the effects of
the top and bottom boundaries are dominant and a modified function 
$f^{\rm ref}(r)=(1-\rho)/(4\rho \pi KR) \times \exp \left( -\rho R/L \right)$
is more adapted  (see Eq.~(\ref{space_cowboy})).
This method yields a very good and rapid convergence as long as
sources are located in the thin disk $z=0$.

                           %*****%

\subsection{Integral representation for infinite radius disk}
\label{integ_repres}

When the disk has an infinite radius, the Bessel sum can be replaced by
an integral, and the end result is obtained from the Bessel sum by
the substitution
$\zeta_i/R \rightarrow k$ and $1/J_1^2(\zeta_i) \approx
\pi \zeta_i/2 \rightarrow k \pi R/2$, so that in the general case
-- see Eq.~(\ref{sol_prim}) and Eqs.~(\ref{sol_gen}) --,
\begin{equation}
    N^L(r, z) =\exp\left(
    \frac{-V_cz}{2K}
    \right)
    \int_0^\infty \frac{k J_0(kr)}{2RA(k)}
    \frac{\sinh\left\{ S(k)(L-z)/2 \right\}
    }{\sinh\left\{ S(k)L/2 \right\} }
    \, dk  \;,
    \label{expression_integrale_complete}
\end{equation}
with
\begin{equation}
       A(k) = 2h \Gamma_{\rm inel} + V_c + KS(k) \coth \left(
       \frac{S(k)L}{2} \right)
       \;\;\;\mbox{and}\;\;\;
       S^2(k) = \frac{V_c^2}{K^2} + \frac{4\Gamma_{\rm rad}}{K} +4 k^2.
\end{equation}
The integrals of the form
\begin{equation}
        I[f] \equiv  \int_0^\infty J_0(x) \, f(x) dx
\end{equation}
where $f(x)$ is a function such that $f(\infty) = 1$ are quite tricky
to compute numerically, due to the very slow decrease of the
oscillations in the integrand.
Two remarks are of great help. First, as $\int_0^\infty J_0(x) dx
=1$, we have
\begin{equation}
        I[f] =  1 - \int_0^\infty J_0(x) \, (1-f(x)) dx\;.
\end{equation}
The convergence is faster, as $1-f \rightarrow 0$ when $x
\rightarrow \infty$.
Second, using the identity $(xJ_1)' = xJ_0$ and integrating by parts,
one has
\begin{equation}
        I[f]
        = \int_0^\infty J_1(x) \,
        \left( \frac{f(x)}{x} - f'(x)\right) dx\;.
\end{equation}
This expression is meaningful only if $x f(x)$ is
a bounded function near the origin $x=0$.
Using the identity $J_0' = -J_1$ and integrating by parts again,
\begin{equation}
        I[f] =   \left[J_0(x)\left(\frac{f(x)}{x}-f'(x)\right)
        \right]^\infty_0
        +\int_0^\infty J_0(x) \,
        \left(f''(x) - \frac{f'(x)}{x} + \frac{f(x)}{x^2} \right)
        dx\;.
\end{equation}
The latter expression is meaningful only if $x^2 f(x)$ is
a bounded function near the origin $x=0$.
These expressions provide several efficient alternatives 
to evaluate $I[f]$.

%%%%%%%%%%%%%%%%%%%%%%%%%%%%%%%%%%%%%%%%%%%%%%%%%%%%%%%%%%%%%%%%%%%%%%%%%%%%%%%%%%%%
                 %%%%%%%%%%%%%%%%%%%%%%%%%%%%%%%%%%%%%%%%%%%%%%%
                 %%%%%%%%%%%%%%%%%%%%%%%%%%%%%%%%%%%%%%%%%%%%%%%

\section{Alternative description of spallations: random walk approach}
\label{zero_crossing}
     A Cosmic Ray crossing the Galactic disk has a probability $p$ to
     disappear in a nuclear reaction with interstellar matter. This probability is related to the
     reaction cross section $\sigma$ by
     \begin{displaymath}
         p= \kappa_1 \sigma n_{\rm ISM} h = 6  \times 10^{-5} \, \kappa_1 \,
         \frac{\sigma}{100
         \unit{mb}}\;,
     \end{displaymath}
     where $\kappa_1 \sim 1$ contains the dependence on the incidence angle of
     the particle with the Galactic plane.
     The propagation in the $z$ axis is a
     one-dimensional random walk, so that for a Cosmic Ray emitted in the
     disk and reaching again the disk after a number $t^\star$ of
     elementary random steps, the probability distribution of
     disk-crossing numbers $n$ is given by (Papoulis 2002)
     \begin{displaymath}
         d{\cal P} (n | t, z(t) = 0 ) = \frac{2n}{\kappa_2 t^\star} \exp \left(
         - \frac{n^2}{\kappa_2 t^\star} \right) \, dn\;.
     \end{displaymath}
     In this expression, $t^\star$ is the number of steps of the walk $z=\sum_{i=1}^{t^\star} z_i$
     and $\kappa_2 \sim 1$ depends on its statistical properties 
     (for instance, $\kappa_2 = 2$ for elementary steps
     $z_i = \pm \lambda$ and $\kappa_2 \approx 1.43$ for $z_i$ uniformly
     distributed in the interval $[-\lambda,\lambda]$).
     The diffusion coefficient is defined as
     \begin{displaymath}
         K \equiv \frac{\langle z^2 \rangle}{2t}
         = \frac{\kappa_3 \lambda^2 t^\star}{2t}
         = \frac{\kappa_3 \lambda^2}{2\tau} = \frac{1}{2} \, \kappa_3 \tau
	 v^2\;,
     \end{displaymath}
     where $\kappa_3\equiv \langle z_i^2 \rangle/\lambda^2$ is the variance of
     the elementary random step (in units of $\lambda$),
     so that the physical time is related to $t^\star$ by
     $t = t^\star \times \tau = t^\star \times 2K/\kappa_3 v^2$,
     where $\lambda$ is the mean free path and $v$ the velocity.
     We thus finally have,
     \begin{displaymath}
         d{\cal P} (n | t, z(t) = 0 ) = \frac{4Kn}{\kappa_2 \kappa_3 v^2
t} \exp \left(
         - \frac{2Kn^2}{\kappa_2 \kappa_3 v^2 t}\right) \, dn
     \end{displaymath}
     We are now able to compute the probability distribution of disk
     crossings for Cosmic Rays emitted from a distance $r$ in the disk as
     \begin{equation}
         \frac{d{\cal P} (n | r, z(t) = 0 )}{dn} = \int_0^\infty
         d{\cal P} (n | t, z(t) = 0 ) {\cal P} (t | r, z(t) = 0 ) \,dt\;,
         \label{integrale_1}
     \end{equation}
     where the probability that a CR reaching distance $r$ in the disk
     was emitted at time $t$ is
     \begin{displaymath}
         {\cal P} (t | r, z(t) = 0 ) \propto \frac{1}{(Kt)^{3/2}}
         \exp \left( - \frac{r^2}{4K t}\right)\;.
     \end{displaymath}
     The above integral \ref{integrale_1} can be performed, yielding the
final result
     \begin{equation}
         \frac{d{\cal P} (n | r, z(t) = 0 )}{dn} = \frac{r_0^2 n}{r^2}
         \left( \displaystyle 1 + \frac{r_0^2 n^2}{r^2} \right)^{-3/2}\;,
         \label{final_nb_crossing}
     \end{equation}
     with
     $r_0^2 \equiv 8 K^2/\kappa_2 \kappa_3 v^2$.
     The average number of disk crossings is readily obtained:
     \begin{displaymath}
         \langle n \rangle \equiv \int_0^\infty n \frac{d{\cal P} (n | r,
         z(t) = 0 )}{dn} \, dn = \frac{r}{r_0}\;,
     \end{displaymath}
     and the associated variance is tends to infinity.
     We can also compute the integrated probability, that more that
$n_0$ crossings have
     occurred, as
     \begin{displaymath}
         {\cal P} (n>n_0 | r, z(t) = 0 ) =
         \left( \displaystyle 1 + \frac{r_0^2 n_0^2}{r^2}
         \right)^{-3/2} \;.
     \end{displaymath}
     A particle having crossed $n$ times the disk has the probability
     $p_n = (1-p)^n \sim \exp(-np)$ of surviving, so that the survival
     probability at distance $r$ is given by
     \begin{displaymath}
         {\cal P}_{\rm surv} (r) = \int_0^\infty \frac{d{\cal P} (n | r,
z(t) = 0 )}{dn}
         \, e^{-np} \, dn =
         \int_0^\infty \frac{x \, dx}{\left(1+x^2\right) ^{3/2}} \,
         e^{-xrp/r_0}\;.
     \end{displaymath}
     This can be written as
     \begin{displaymath}
         {\cal P}_{\rm surv} (r) =
         \int_0^\infty \frac{x \, dx}{\left(1+x^2\right) ^{3/2}} \,
         e^{-\alpha x r/r_{\rm spal}}\;,
     \end{displaymath}
     with $r_{\rm spal}$ defined in Eq.~(\ref{f_rspal}) and $\alpha \equiv
     \kappa_1 \sqrt{\kappa_2 \kappa_3/2}$.
The density of Cosmic Rays in the disk is then given by
\begin{equation}
      N(r) = \frac{1}{4\pi K r} \times  {\cal P}_{\rm surv} (r)
      = \frac{1}{4\pi K r} \int_0^\infty \frac{x \,
      dx}{\left(1+x^2\right) ^{3/2}} \,
      e^{-\alpha x r/r_{\rm spal}}\;.
      \label{The_riri_equation}
\end{equation}
The quantity $\alpha$ seems to be related to the detailed statistical
properties of the random walk under consideration, through
$\kappa_1$, $\kappa_2$ and $ \kappa_3$. However, direct
comparison with the alternative
expression~(\ref{forme_utile})
obtained above indicates that these expressions are indeed
equivalent, with $\alpha = 1$.

%%%%%%%%%%%%%%%%%%%%%%%%%%%%%%%%%%%%%%%%%%%%%%%%%%%%%%%%%%%%%%%%%%%%%%%%%%%%%%%%%%%%
%%%%%%%%%%%%%%%%%%%%%%%%%%%%%%%%%%%%%%%%%%%%%%%%%%%%%%%%%%%%%%%%%%%%%%%%%%%%%%%%%%%%

\bibliographystyle{aa}
\bibliography{disk_final}

\end{document}